\documentclass[final,5p,times,twocolumn]{elsarticle}

\usepackage[utf8]{inputenc}
\usepackage[T1]{fontenc}

\graphicspath{{graphics/}}

\DeclareGraphicsExtensions{.pdf,.jpeg,.png}

\usepackage{amsmath}
\usepackage{amsfonts}
\usepackage{amssymb}
\usepackage{float}

\usepackage[normalem]{ulem}

\usepackage{booktabs}
\usepackage{tabularx}
\usepackage{threeparttable}
\usepackage{siunitx}

\usepackage{url}
\usepackage[colorlinks=true, citecolor=blue, linkcolor=blue, filecolor=blue,urlcolor=blue]{hyperref}

\usepackage[gen]{eurosym}

\def\el{${}_{\textrm{el}}$}
\def\th{${}_{\textrm{th}}$}

\newcommand{\ubar}[1]{\text{\b{$#1$}}}

\def\l{\lambda}

\def\m{\mu}
\def\G{\Gamma}
\def\g{\gamma}
\def\d{\partial}
\def\cL{\mathcal{L}}

\usepackage{tikz}

\usepackage[europeanresistors,americaninductors]{circuitikz}
\usepackage{adjustbox}

\usepackage{fixltx2e}
\journal{Energy Economics}

\begin{document}
%\linenumbers
\begin{frontmatter}



\title{Decreasing market value of variable renewables can be avoided by policy action}


\author[tub,kit]{T.~Brown\corref{cor1}}
\ead{t.brown@tu-berlin.de}
\author[chalmers,aalto]{L.~Reichenberg}

\cortext[cor1]{Corresponding author}
\address[tub]{Department of Digital Transformation in Energy Systems, Technische Universität Berlin, Einsteinufer 25 (TA 8), 10587 Berlin, Germany}
\address[kit]{Institute for Automation and Applied Informatics, Karlsruhe Institute of Technology, Hermann-von-Helmholtz-Platz 1, 76344 Eggenstein-Leopoldshafen, Germany}
\address[chalmers]{Department of Space Earth and Environment, Chalmers University of Technology, 412 96, Göteborg, Sweden}
\address[aalto]{Department of Mathematics and Systems Analysis, Aalto University, Otakaari 1 F, Espoo, Finland}

\begin{abstract}
  %\boldmath
Although recent studies have shown that electricity systems with shares of wind and solar above 80\% can be affordable, economists have raised
concerns about market integration. Correlated generation from variable renewable sources depresses market prices, which can cause wind and solar to cannibalise their own revenues and prevent them from covering their costs from the market. This cannibalisation appears to set limits on the integration of wind and solar, and thus to contradict studies that show that high shares are cost effective. Here we show from theory and with
simulation examples how market incentives interact with prices, revenue and
costs for renewable electricity systems.  The
decline in average revenue seen in some recent literature is due to an implicit policy assumption that technologies are forced into the system, whether it be with
subsidies or quotas. This decline is mathematically guaranteed regardless of whether the subsidised technology is variable or not.
If instead the driving policy is a carbon dioxide cap or
tax, wind and solar shares can rise without cannibalising their own market
revenue,  even at
penetrations of wind and solar above 80\%.
The strong dependence of market value on the policy regime means that market value needs to be used with caution as a measure of market integration.
Declining market value is not necessarily a sign of integration problems, but rather a result of policy choices.

\end{abstract}

\begin{keyword}
%Here are some suggestions:
 market value of variable renewables \sep renewable energy policy \sep CO$_2$ tax \sep feed-in premium \sep merit order effect \sep large-scale integration of renewable power generation



\end{keyword}

\end{frontmatter}

\section{Highlights}

\begin{itemize}
  \item Decreasing market value (MV) with wind and solar share can be avoided by CO$_2$ pricing
  \item In long-term equilibria, wind and solar subsidies reduce MV, but CO$_2$ prices do not
  \item Models with rising CO$_2$ prices draw in wind and solar (VRE) with no reduction in MV
  \item Falling MV in models with VRE subsidy does not necessarily indicate integration problems
  \item These results are confirmed using theory and in a power system model
\end{itemize}

\section{Introduction}

Rising shares of wind and solar in electricity markets around the world have led to concerns about their market integration at high penetrations.
Several studies have found empirical evidence that electricity prices have decreased in markets as the share of variable renewable energy (VRE) has risen \cite{Sensfuss2007_1000007777,SENSFU20083086,7080941,Figueiredo2017, ozdemir2017integration,Hirth2018143,LopezProl2020,MILLS2015269}. The cause of the lower prices is the very low or zero marginal cost of wind and solar generators. This pushes out some of the more expensive generators from the market, and, since the price is usually set by the marginal cost of the last generator needed to satisfy demand, the prices are depressed during times of wind and solar generation. Lower prices lead to lower revenues for all generators (the `merit order effect' \cite{SENSFU20083086}), but especially so for wind and solar generators, since their generation depresses prices exactly when they are generating most, an effect known as `cannibalisation' \cite{LopezProl2020,MILLS2015269}.
Both the generally lower prices and the cannibalisation effect have been perceived as problematic, because they lead to lower market revenues and would lead to less incentive to invest in new capacity in a free market \cite{JOSKOW2008159,7080941}.

These empirical observations \cite{Sensfuss2007_1000007777,SENSFU20083086,7080941,Figueiredo2017, ozdemir2017integration,Hirth2018143,LopezProl2020,MILLS2015269} were made in electricity systems where the existing conventional power generation fleet remained largely unchanged, i.e. in the short-term.
While short-term effects are important,  not least because they are presently faced by actors on the market, the long-term effects, i.e. the situation after the capacity mix has adjusted to an equilibrium state, set a limit on the \emph{possible} role that VRE may play in the power system. For this reason, the long-term effects are the focus of this paper.
Using computer models of the power market where investments in all generator capacities are optimised, it has been shown that even in a long-term equilibrium there is a decline in revenue for wind and solar with their penetration \cite{LAMONT20081208,Joskow2011,kopp2012,mills2013,HIRTH2013218,gilmore2014,MILLS2015269,WINKLER2016464,gowrisankaran2016}.
The size of the effect was estimated to be a decrease in revenues by half at penetration levels of 15\% for solar and 30\% for wind in a selection of European countries \cite{HIRTH2013218}. The hypothesis put forward in these papers is that it is the variability of wind and solar that causes the decline \cite{LAMONT20081208,HIRTH2013218,WINKLER2016464,Hirth2016}.

The idea that variability sets a ceiling for the cost-effective penetration level of VRE electricity has been influential.
%Hirth concludes that the long-term market value decreases by nearly half for penetration levels beyond ~30\% (wind) and ~15\% (solar).
Blazquez \cite{BLAZQUEZ20181} claims that `The Paradox holds as long as market clear prices with short term marginal costs, and renewable technology's marginal cost is close to zero and not dispatchable', i.e. that energy-only markets with variable renewables inherently entail decreasing market value with penetration levels. Some have even suggested that the capacity factor (typically 10-25\% for solar and 20-40\% for onshore wind) should be considered as a limit on penetration \cite{JenkinsTrembath2015}. Hirth and Radebach \cite{Hirth2016} conclude: `Finally, and more fundamentally, it [the declining MV of VRE with penetration level] indicates that variable renewables face a substantial difficulty in becoming economical at high market shares. Without fundamental technological breakthroughs, a deep decarbonisation of power systems will be hard to achieve based on wind and solar power alone. Other supplementary low carbon technologies are likely to be needed.'

In apparent contradiction to the above-mentioned market integration studies,
the last few years have also seen an increasing number of cost-minimising energy system studies with high shares (>80\%) of variable renewables \cite{burdenresponse,HANSEN2019471,fripp2012,HALLER2012282,budischak2013,Breyer_2015,BOGDANOV2016176,FREW201665,Schlachtberger2017,Brown2018}. The system solutions of these studies correspond to long-term equilibria where all generators, including VRE technologies, exactly cover their costs with their market revenue (the `zero-profit rule' \cite{boiteux1949,boiteux1960}). This appears to contradict market value studies that claim that wind and solar revenue will be pushed below the cost-recovery level at high penetrations.

We resolve this contradiction by showing in theory and in model simulations how cost, revenue and policy interact. We show that market value studies find declining market value by construction because they choose to force in VRE with support policies (be it quotas, feed-in tariffs, feed-in premiums or capacity incentives). If instead rising carbon dioxide (CO$_2$) taxes are used as the primary policy instrument to draw in wind and solar, then VRE revenue will always be sufficient to cover generation costs. We demonstrate that this holds in a power system model even at a penetration of solar and wind above 80\%, which is much higher than is usually considered in the market value literature.

While the effects on market prices of subsidies for desirable goods versus taxes on undesirable goods is well understood \cite{pigou1920,baumol1972}, the effects on market values for different technologies in long-term electricity market equilibria has not been sufficiently considered in the literature.
Many market value studies have considered the impact of fixed CO$_2$ taxes \cite{kopp2012,mills2013,HIRTH2013218,gilmore2014,hirth2014,gowrisankaran2016,WINKLER2016464,ZIPP20171111}, but CO$_2$ taxes have only ever been a subordinate policy to the main policy of VRE support. The resolution of the contradiction requires replacing VRE support with CO$_2$ reduction incentives. Although the purpose of a CO$_2$ tax is not explicitly to increase VRE generation, but instead to reduce CO$_2$ emissions up to the point where the marginal abatement cost is equal to the tax, it will have the indirect effect of drawing in VRE generators if they are the most cost-effective low-emission generators available in the system.

We show how the market value of variable renewables, and indeed any type of generator regardless of its variability, is contingent on the mechanism (policy instrument) by which it enters the long-term equilibrium solution. We also discuss how the previous literature on the subject has failed to highlight this fact, either by an implicit assumption on the policy instrument \cite{kopp2012,mills2013,HIRTH2013218,gilmore2014,MILLS2015269,WINKLER2016464,gowrisankaran2016}, by simultaneously changing several confounding factors \cite{HIRTH2013218,WINKLER2016464} or by ignoring market value altogether \cite{Schlachtberger2017,REICHENBERG2018914, FREW201665}.

First we consider the economic theory (Sections \ref{sec:economic_theory} and \ref{sec:math_theory}) and then demonstrate the effects in a reimplementation  of the energy system model EMMA used in \cite{HIRTH2013218} in the PyPSA modelling framework \cite{PyPSA} (Sections \ref{sec:numericalmodel} and \ref{sec:numerical_results}).

\section{Introduction to theory}
\label{sec:economic_theory}

\subsection{Zero-profit rule, market value and LCOE without policy measures}\label{sec:equ}

In a long-term equilibrium, where generator capacity is optimised along with power system operation, producers make zero profit under idealised conditions of perfect market competition, free entry and exit, linear cost functions and without any further constraints \cite{boiteux1949,boiteux1960}. If any producer makes a net profit, new producers will enter the market and competition will drive profits to zero; similarly if producers make a net loss, some will exit the market until losses are eliminated. For electricity markets, this \emph{zero-profit condition} means in a long-term equilibrium that the average revenue that generators receive from the market exactly covers their costs.

The zero-profit condition can be restated per unit of generated energy in terms of the market value $MV_s$ and levelised cost of electricity $LCOE_s$ of each generator $s$:
\begin{equation}
    MV_s = LCOE_s  \label{eq:main}
\end{equation}
The market value is defined as the revenue averaged over each unit of energy sold. The LCOE is defined as the net present sum of all investment, fuel, operational and maintenance costs averaged over each unit of energy that is actually generated. While this definition of LCOE agrees with the usual definition for dispatchable generators, we differ from the standard definition for wind and solar by only averaging over the actual energy generated, rather than what theoretically would be available before curtailment. This definition raises the LCOE of wind and solar when there is curtailment at high penetrations.  The equality \eqref{eq:main} is proved for a general long-term equilibrium power model in Section \ref{sec:no_network}.

Different generators have different market values because they occupy different niches in the optimal system, depending on their characteristics such as cost and variability. Each technology has its own optimal share of generation in the long-term equilibrium. To change that share, policy intervention is required. In this contribution we consider both support policies that force particular technologies into the system, as well as policies that force out polluting technologies.

\subsection{Technology-specific support policies}

Under `support policy' we group all policies that encourage investment in a particular technology beyond the pure cost-optimum, either by mandating a certain share in the generation mix, or by creating a revenue stream independent of the electricity market. Examples of such policies include Renewable Energy Portfolio Standards (RPS) in various US states and Feed-in Tariffs (FiT) in Germany, which give a remuneration for every unit of energy generated from wind and solar.

The fact that an additional subsidy is required to achieve a higher share implies that the generator cannot make sufficient revenue from the market to cover its costs. The subsidy required to cover costs can be translated into an equivalent Feed-in Premium (FiP) $\mu_s > 0$ paid per unit of generated energy, thus modifying the zero-profit condition at equilibrium to
\begin{equation}
    MV_s = LCOE_s - \mu_s  \label{eq:main:vre}
\end{equation}
The FiP tops up the average revenue received by the generator from the market to the LCOE so that the generator covers its costs.

This relation holds regardless of the technology or support policy. If a technology is forced to cover a fixed share of demand, we show in Section \ref{sec:retproof} that $\mu_s$ is the shadow price of the corresponding constraint. If the share of generation available before curtailment is fixed instead of the actual generation \cite{HIRTH2013218,pahle2015}, $\mu_s$ is proportional to the shadow price of the constraint (see \ref{sec:retcapproof}). If generators do not participate in the market at all, but are paid a Feed-in Tariff (FiT) to cover their costs at the same level as the LCOE, then $\mu_s$ is the difference between the average market value and the tariff.

Without a support policy, the system would settle into the equilibrium described in Section \ref{sec:equ} with MV equal to LCOE for all technologies. A higher share of technology $s$ requires a policy intervention. The higher the share, the higher the equivalent FiP $\mu_s$ needs to be, and thus the lower the market value drops according to equation \eqref{eq:main:vre}.

The decline in market value is an indirect effect of the decline of market prices during the hours that the supported technology is generating. The exact mechanism is explained in more detail in the next section, but essentially the FiP $\mu_s$ paid per MWh reduces the effective marginal cost of technology $s$ by $\mu_s$, since it gets a revenue of $\mu_s$ from outside the market. The combined effect of having a larger share of technology $s$ as well as technology $s$ bidding into the market at a lower price serves to lower market prices when $s$ is generating. This well-known fact that prices are surpressed by a surplus of some good is the essential result observed in the literature on the market value of renewables \cite{borenstein2008,kopp2012,mills2013,HIRTH2013218,gilmore2014,MILLS2015269,gowrisankaran2016,WINKLER2016464}.
Yet, these studies do not draw the conclusion that the MV decline is due to the surplus of VRE, but instead explain it by the variable nature of these sources.

The MV decline does not contradict the zero-profit condition, since the condition only applies to an undistorted equilibrium. We have departed from the equilibrium solution by forcing a share of a technology. The other technologies are still freely optimised, and are thus still subject to the zero-profit rule, although their share of total generation will be lower.

\subsection{CO$_2$ policies}

CO$_2$ policies include direct CO$_2$ taxes and CO$_2$ caps with traded certificates. They indirectly support wind, solar and other low-emission technologies by penalising high-emission generators.

%CO$_2$ policies alter the zero-profit rule %according to the equivalent CO$_2$ tax
Under CO$_2$ policies, the zero-profit rule still holds, but the relationship between revenue and costs now includes the equivalent CO$_2$ tax
$\mu_{\textrm{CO}_2}$ (in euro per tonne of CO$_2$, \euro tCO$_2^{-1}$) and the technology-specific emission factor $e_s$ (in tCO$_2$MWh$^{-1}$)
\begin{equation}
    MV_s = LCOE_s + e_s\mu_{\textrm{CO}_2} \label{eq:main:co2}
\end{equation}
This relation is proven in Section \ref{sec:co2proof}.

For technologies like wind, solar or nuclear with no direct emissions, we have exact cost recovery $MV_s = LCOE_s$. CO$_2$-emitting generators have to cover both generation costs and the CO$_2$ tax with their market revenue at equilibrium, and are thus pressured out of the market to the benefit of low-CO$_2$ generation.

CO$_2$ policies raise the market values of CO$_2$-emitting generation by raising their marginal costs, and thus raising prices at hours when they are generating.

\subsection{Comparison of support and CO$_2$ policies}

\begin{figure}[!t]
\centering
%graphic generated in mv_example-200125-112718.ipynb
% left bottom right top
    \includegraphics[trim=2cm 2cm 2.4cm 1cm,width=\linewidth,clip=true]{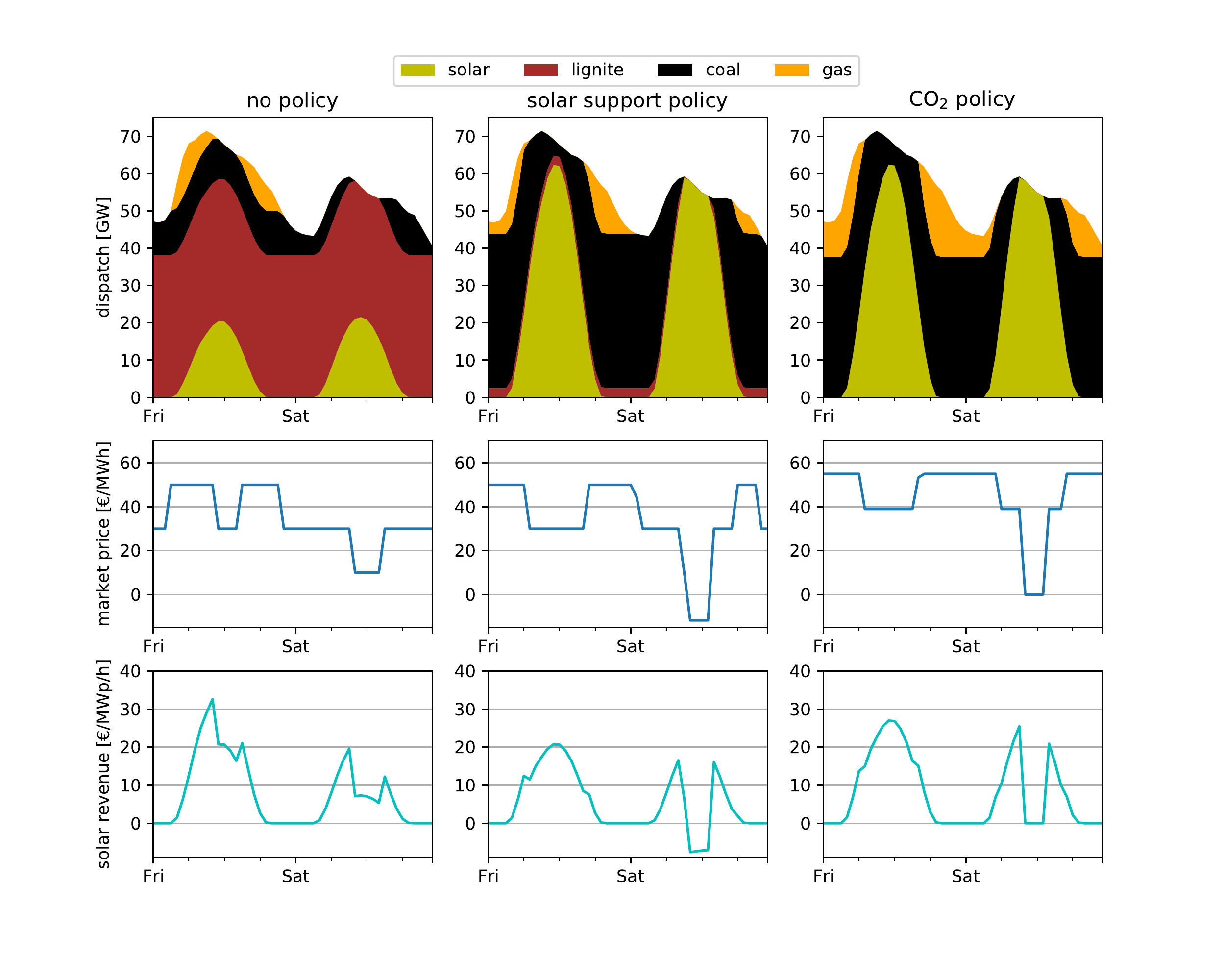}
\caption{Comparison of dispatch, market price and momentary solar revenue per unit of capacity in a highly-simplified long-term equilibrium model over two days with only four technologies (lignite, coal, gas and solar) and (a) no policy, (b) solar support policy and (c) CO$_2$ policy. The solar support and CO$_2$ policies are tuned to given the same penetration of solar.}
\label{fig:comparison-example}
\end{figure}

The effects of the two types of policy on prices and market values are strikingly different.  Support policies depress market prices when the supported generators are running and offer them compensation outside of the market, whereas CO$_2$ policies raise market prices when fossil-fuelled generators are running, thus encouraging low-emission generators into the market. Support policies increase the share of low-emission technologies but reduce their average market revenue, whereas CO$_2$ policies increase their share while leaving their zero-profit condition intact. For fossil-fuelled generators, low-carbon support policies reduce their share of the market but do not affect their zero-profit condition, while CO$_2$ policies increase the overall costs they need to cover from the market, thus also reducing their share.

Figure \ref{fig:comparison-example} provides an illustration of how the two policies impact dispatch, price and momentary revenue in a highly-simplified model with solar and three fossil technologies (lignite, hard coal and gas) over a period of two days (more elaborate simulations are provided in later sections). The solar support policy lowers prices both by the merit order effect when solar is feeding in and by turning prices negative when solar is price-setting (see the next section for a discussion of this mechanism; solar generation continues during this period because its feed-in is being subsidised outside of the market). Under the CO$_2$ policy, prices go to zero when solar is price-setting, but this is more than compensated by the rise in prices when the fossil-fuelled generators are price-setting.

These effects on prices affect the market value of solar. With no policy, there is sufficient revenue for solar to cover its costs. For the CO$_2$ policy this is also the case, since the area under the revenue curve is the same, but the hours when solar earns change: it earns less at midday, but more on the flanks of its generation profile. For the solar support policy, prices and revenue are lower at all times when it generates, so solar cannot cover its costs from the market.

Additional flexibility options such as transmission, demand response and storage alter the background system by allowing price arbitrage to smooth the variability of renewable generators \cite{HIRTH2013218,HIRTH2016210,Tveten2016}. By providing more demand in hours with low prices, flexibility helps to raise prices when renewables are abundant \cite{Brown2019,haertel2021,bernath2021,boettger2021,Ruhnau2020Market}. When renewables are scarce and prices are high, flexibility lowers prices by providing more supply. More flexibility means that lower subsidies are required for VRE support policies to reach a given penetration level, while for CO$_2$ policies a lower CO$_2$ price is required for a given abatement level when flexibility is available.

\section{Theory}
\label{sec:math_theory}

In this section we use a long-term optimisation framework to show how prices and market values relate to costs and policy measures, and in particular under what circumstances the `zero profit' rule holds. Proofs are provided for the equations stated in the previous section. The proof of the zero-profit condition with no additional policy measures goes back to Boiteux (1949) \cite{boiteux1949}; the discussion of profit under VRE support policies can also be found in Green \& Léautier (2015) \cite{gl2015}. The optimisation problem setup and use of Karush-Kuhn-Tucker (KKT) conditions in the present contribution follows the textbook by Biggar and Hesamzadeh \cite{2014}.

\subsection{Long-term equilibrium without policy measures}\label{sec:no_network}

We maximise yearly social welfare for a single node with linear supply cost functions in a long-term equilibrium:
\begin{equation}
    \max_{d_{a,t}, g_{s,t}, G_s}\left[\sum_{a,t} U_{a,t}(d_{a,t}) -  \sum_s c_s G_s - \sum_{s,t} o_{s} g_{s,t}\right]  \label{eq:obj}
\end{equation}
subject to
\begin{align}
   \sum_a d_{a,t} - \sum_s g_{s,t} & =  0 \hspace{0.34cm}\perp \hspace{0.34cm} \l_t \hspace{0.34cm} \forall t \label{eq:balance}\\
    -g_{s,t} & \leq 0 \hspace{0.34cm}\perp \hspace{0.34cm} \ubar{\mu}_{s,t} \hspace{0.34cm} \forall s,t  \label{eq:inequ1}\\
         g_{s,t} - \bar{g}_{s,t} G_s & \leq 0 \hspace{0.34cm}\perp \hspace{0.34cm} \bar{\mu}_{s,t} \hspace{0.34cm} \forall s,t \label{eq:inequ2}
\end{align}
Here $t$ labels time periods representing a year of load and weather conditions, $a$ labels consumers, $s$ labels generators, $d_{a,t}$ is the
demand, $g_{s,t}$ is the generator dispatch, $G_s$ is the generator
capacity and $\bar{g}_{s,t}\in[0,1]$ is the availability/capacity factor (which
varies with time for variable renewable generators like wind and
solar). $\l_t$ is the marginal price of electricity, while
$\bar{\mu}_{s,t}$ and $\ubar{\mu}_{s,t}$ represent shadow prices of
the generator constraints. $c_s$ represent the annualised investment and fixed operations and maintenance costs of the generators, while
$o_s$ represent variable costs. $U_{a,t}(d_{a,t})$ are the differentiable, concave utility
functions of the consumers.

The KKT conditions are first-order conditions that are necessary
for the optimal solution to satisfy. (Since in our case the
objective function is concave and the constraints are affine, the conditions are also sufficient for optimality). The
definition of the Lagrangian $\cL$ and KKT conditions are provided in
\ref{sec:kkt}. Conventions are chosen such that $\l_t$ is positive if
the price-setting generator has positive marginal costs, and such that
all shadow prices $\m$ are positive or zero.

From KKT stationarity we have for the variables representing the generator dispatch $g_{s,t}$ and capacity $G_s$:
\begin{align}
    \frac{\d \cL}{\d g_{s,t}} = 0 &\Rightarrow  -o_{s}+\l_t + \ubar{\m}_{s,t} - \bar{\m}_{s,t}  = 0  \label{eq:statvanilla}\\
    \frac{\d \cL}{\d G_{s}} = 0 &\Rightarrow  -c_s + \sum_t \bar{g}_{s,t}\bar{\m}_{s,t}  = 0  \label{eq:statcapvanilla}
\end{align}
while for the inequalities  \eqref{eq:inequ1} and  \eqref{eq:inequ2} we get from KKT complementary slackness:
\begin{align}
  \ubar{\m}_{s,t} g_{s,t} & = 0\\
  \bar{\m}_{s,t} (\bar{g}_{s,t}G_s-g_{s,t}) & = 0
\end{align}

We will now show that each generator $s$ exactly makes back their costs $c_s G_s + \sum_{t} o_s g_{s,t}$ from their market revenue $\sum_t \l_t g_{s,t}$, i.e. the `zero-profit condition'.
\begin{align}
  c_s G_s + \sum_{t} o_{s} g_{s,t} & = \left( \sum_t  \bar{g}_{s,t}\bar{\m}_{s,t}\right) G_s + \sum_{t} \left(\l_t + \ubar{\m}_{s,t} - \bar{\m}_{s,t}\right) g_{s,t} \nonumber \\
  & = \sum_{t}\left[\l_t g_{s,t} + \bar{\m}_{s,t}(\bar{g}_{s,t} G_s- g_{s,t}) + \ubar{\m}_{s,t} g_{s,t}\right]\nonumber \\
  & = \sum_{t}\l_t g_{s,t} \label{eq:noprofit-vanilla}
\end{align}
The first step substitutes the equations from KKT stationarity; in the second step terms are reorganised; in the final step the equations from KKT complementary slackness are applied.

We can use this, along with primal feasibility for the demand balancing constraint \eqref{eq:balance}, to show that the total generator costs are equal to the total payments by consumers:
\begin{align}
  \sum_s \left[ c_s G_s + \sum_{t} o_{s} g_{s,t}\right]
   = \sum_{s,t}\l_t g_{s,t}
   = \sum_{a,t} \l_t d_{a,t} \label{sec:total_costs}
\end{align}

For a situation with perfectly price-inelastic demand where we can reduce the overall problem to generator cost minimisation, this is the statement of strong duality between the objectives of the primal and dual problems. Note, however, that unlike equation \eqref{sec:total_costs}, equation \eqref{eq:noprofit-vanilla} also holds at the level of individual generators.

\subsection{LCOE, MV and RMV without policy measures}

When both sides of \eqref{eq:noprofit-vanilla} are divided by the generator's total dispatch we recover on the left the definition of the levelised cost of energy (LCOE)  of the generator:
\begin{equation}
  LCOE_s \equiv \frac{c_s G_s + \sum_t o_s g_{s,t}}{\sum_t g_{s,t}}
\end{equation}
and on the right the definition of the market value (MV) of the generator, sometimes called the absolute market value \cite{HIRTH2013218}, which gives us the average revenue when the generator is producing:
\begin{equation}
  MV_s \equiv \frac{\sum_t g_{s,t} \l_t}{\sum_t g_{s,t}}
\end{equation}
The equality \eqref{eq:noprofit-vanilla} then gives us:
\begin{equation}
  LCOE_s = MV_s  \hspace{0.34cm} \forall s
\end{equation}
This is a restatement of the zero-profit rule on an averaged per-MWh basis.

The relative market value (RMV), also called the value factor in \cite{HIRTH2013218}, is the ratio of the market value to the load-weighted average market price:
\begin{equation}
  RMV_s \equiv MV_s \left(\frac{\sum_{a,t} d_{a,t}\l_t }{\sum_t d_{a,t}} \right)^{-1} =  \frac{\left(\sum_t g_{s,t} \l_t \right)\left(\sum_{a,t} d_{a,t}\right)}{\left(\sum_t g_{s,t}\right)\left(\sum_{a,t} d_{a,t} \l_t\right) } \label{eq:rmvdef}
\end{equation}

Using the zero-profit rule \eqref{eq:noprofit-vanilla} and the energy balance constraint \eqref{eq:balance}  we can rewrite the RMV:
\begin{equation}
  RMV_s = \left( \frac{c_s G_s + \sum_t o_s g_{s,t}}{\sum_r c_r G_r + \sum_{r,t} o_r g_{r,t}} \right)\left(\frac{\sum_{t} g_{s,t}}{\sum_{a,t} d_{a,t}}\right)^{-1} \label{eq:rmvid}
\end{equation}
From this it can be seen that in the absence of other constraints, the RMV is the ratio of a technology's share of total costs (first fraction) to its share of demand (second fraction). If a particular technology has a similar share of both energy provision and costs, then it will have an RMV close to unity.

\subsection{Long-term equilibrium with support policy}\label{sec:retproof}

If a subset of generators $S$ is singled out and forced to provide a fixed amount of energy $\Gamma$ during the year, this is represented with the constraint
\begin{equation}
  \sum_{s\in S,t} g_{s,t} \geq \G \hspace{0.34cm}\perp \hspace{0.34cm} \m_{\G} \label{eq:vre}
\end{equation}
For example, for a particular penetration of wind, $S$ would
represent all wind generators and $\G$ would be a fixed fraction of
the annual demand.

For generators included in the constraint, $s\in S$, the stationarity equation \eqref{eq:statvanilla} for $g_{s,t}$ from the previous section is altered to
\begin{align}
    \frac{\d \cL}{\d g_{s,t}} = 0 &\Rightarrow  -o_{s}+\l_t + \ubar{\m}_{s,t} - \bar{\m}_{s,t} + \m_{\Gamma}  = 0
\end{align}
so that now for the generators in $S$
\begin{equation}
  c_s G_s + \sum_{t} o_s g_{s,t} = \sum_t g_{s,t} (\l_t + \m_\G)  \hspace{0.34cm} \forall s \in S \label{eq:noprofit-vre}
\end{equation}

For generators excluded from the constraint, $s \notin S$, the
zero-profit rule remains exactly the same as
\eqref{eq:noprofit-vanilla}.

If \eqref{eq:vre} is not binding, then $\m_\G = 0$ and the zero-profit rule is recovered. In this case the given share is already part of the unconstrained optimum. However if \eqref{eq:vre} is binding, then more generation from $S$ is being forced into the solution than the optimum without constraint \eqref{eq:vre}, therefore $\m_\G > 0$ and generators in $S$ can no longer recover their costs from the market prices $\l_t$ alone. $\m_\G$ represents the per-MWh subsidy, or Feed-in Premium (FiP), required beyond the market price for generators  in $S$ to recover their costs.

Dividing by the total generation $\sum_t g_{s,t}$ we find for $s \in S$
\begin{equation}
  LCOE_s = MV_s  + \m_\Gamma \hspace{0.34cm} \forall s \in S
  \label{eq:ret-relation}
\end{equation}
For $s \notin S$ we have the regular no-profit rule
\begin{equation}
  LCOE_s = MV_s  \hspace{0.34cm} \forall s \notin S
\end{equation}

Expressed another way: forcing in the penetration of a particular
technology above its unconstrained optimal share depresses the market
prices $\l_t$ at the times when it is generating. This accounts for the `market value' effect in
long-term equilibrium models observed in \cite{HIRTH2013218}.

The prices found here can be reproduced by taking the optimal value of $\m_\G$, removing the constraint \eqref{eq:vre} and making the substitution $o_s \to o_s - \m_\G$ for $s\in S$ to get a new, lower effective marginal cost, i.e. moving $\m_\G g_{s,t}$ to the left-hand side of \eqref{eq:noprofit-vre} (see proof in \ref{sec:equivalence}). The support policy thus depresses market prices by two mechanisms: when technology $s$ is generating, the larger share of technology $s$ pushes down prices even when technology $s$ is not price-setting by pushing the supply curve to the right (the merit order effect); when technology $s$ is price-setting, the subsidy reduces the bid $o_s$ by $\m_\G$ and  can even turn the market price negative if
$\m_\G$ is larger than the marginal cost $o_s$. Negative bids are rational for generators if they are guaranteed the subsidy even when prices are negative. In reality, some markets suspend support for subsidised generators bidding in the market once market prices turn negative for a sufficient time (4 hours in the case of renewable energy in Germany built from 2021 \cite{EEG2021}), thereby removing the incentive for them to bid negative prices and thus mitigating this effect. In this case, examined in \ref{app:ret-zero}, support policies still depress prices by the merit order effect.

%Note that if \eqref{eq:vre} had been an equality constraint with
%shadow price $\l_\G$, the behaviour and equations would have been
%exactly the same with the substitution of $\l_\G$ for $\m_\G$, except
%that when $\G$ were below the unconstrained optimal share of generators
%$S$, $\l_\G$ would have turned negative.

\subsection{Long-term equilibrium with CO$_2$ policy}\label{sec:co2proof}

If, rather than supporting particular technologies, we replace
constraint \eqref{eq:vre} with a CO$_2$ cap $K$, the behaviour is
different. Consider the CO$_2$ constraint:
\begin{equation}
  \sum_{s,t} e_s g_{s,t} \leq K \hspace{0.34cm}\perp \hspace{0.34cm} \mu_{\textrm{CO}_2} \label{eq:co2}
\end{equation}
where $e_s$ is the emission factor in tonne-CO$_2$ per MWh\el{} for
generator $s$ and $K$ is a cap on yearly emissions in tonne-CO$_2$ per
year. This constraint has the same form as \eqref{eq:vre}, except for the direction of the inequality sign and the weighting of generation.

The stationarity equation \eqref{eq:statvanilla} is altered to
\begin{align}
    \frac{\d \cL}{\d g_{s,t}} = 0 &\Rightarrow  -o_{s}+\l_t + \ubar{\m}_{s,t} - \bar{\m}_{s,t} - e_s\mu_{\textrm{CO}_2}   = 0
\end{align}
and now
\begin{equation}
  c_s G_s + \sum_{t} o_s g_{s,t} = \sum_t g_{s,t} (\l_t - e_s\mu_{\textrm{CO}_2})  \hspace{0.34cm} \forall s \label{eq:noprofit-co2}
\end{equation}
If the constraint \eqref{eq:co2} is binding, it pushes up market
prices beyond the cost-recovery point so that charges for CO$_2$
emissions are also covered from the market.

Dividing by the total generation $\sum_t g_{s,t}$ we find
\begin{equation}
  LCOE_s = MV_s  - e_s\mu_{\textrm{CO}_2} \hspace{0.34cm} \forall s \in S
\end{equation}

In this case, generators with no direct emissions, $e_s = 0$, continue
to satisfy the zero-profit rule. Emitting generators with $e_s>0$ have
to cover the CO$_2$ price with their market revenues, but still
recover their costs once the CO$_2$ levy has been paid.

The same prices can be obtained by replacing the CO$_2$ constraint with a direct cost of CO$_2$ and making the substitution $o_s \to o_s + e_s \mu_{\textrm{CO}_2}$, i.e. moving the term $ e_s\mu_{\textrm{CO}_2} g_{s,t} $ to the left-hand side of  \eqref{eq:noprofit-co2}. Through the higher effective operating costs for CO$_2$-emitting generators, the CO$_2$ price increases market prices when these generators are setting the price.

\subsection{Other setups}

Hybrid setups that combine CO$_2$ pricing and technology support are of course possible, and can be found in many of today's markets in Europe. A moderate CO$_2$ price tilts the equilibrium in favour of low-carbon technologies and reduces the feed-in premium needed for a given share of variable renewable energy, thus raising the market value. For a given penetration of wind and solar, a hybrid approach allows the market value of VRE to be set at any value between the two extremes of low MV with a support policy only, and MV equal to LCOE for the case that only a CO$_2$ policy is used to induce the share of VRE.  Combining CO$_2$ pricing and technology support can reduce market distortions while limiting investor risk and thus their financing costs, as discussed in Section \ref{sec:policy}.

More complicated setups (forcing fixed shares for available rather than dispatched energy, limited installation potentials, multi-node networks, storage, convex generation costs) do not alter the conclusions reached for the simpler model above. Proofs for these setups
can be found in \ref{sec:furtherproofs}.

In addition, we show in \ref{sec:lift} that subsidising a set of technologies is exactly equivalent to taxing all other technologies when demand is perfectly price-inelastic. Switching from subsidy to tax just results in a constant lift to all the prices, and therefore a constant lift to all market values. Since a tax on non-VRE technologies is not a realistic policy proposal, we focus in the main text on CO$_2$ policies.

\section{Power system model description}\label{sec:numericalmodel}

The theoretical insights developed above are demonstrated in a market model based on EMMA \cite{HIRTH2013218} that has been reimplemented in the open PyPSA framework \cite{PyPSA}. The code for the model is available online under an open licence \cite{mvcode}.

The model has five nodes for Germany and four of its neighbours: Poland, France, the Netherlands and Belgium. The model minimises long-term generation costs over historical hourly load and weather from the year 2010, assuming a perfectly price-inelastic demand up until a high value of lost load (1000~\euro/MWh). The model completely rebuilds the existing generation system (`greenfield investment') except for pumped hydro storage, for which existing capacities are taken assuming an energy storage capacity of eight hours at nominal power.

Generators are aggregated into a single representative class for each technology following \cite{HIRTH2013218}. The available variable renewable technologies are wind and solar power, while the dispatchable generators are coal, lignite, lignite with CCS, nuclear, open cycle gas turbines (OCGT) and combined cycle gas turbines (CCGT). We keep most of EMMA's cost and other technical assumptions, but update the nuclear cost from 4000~\euro/kW to 6000~\euro/kW, reflecting recent experience in Europe \cite{schroeder2013}, as well as reducing the wind cost from 1300~\euro/kW to 1040~\euro/kW and the solar cost from 2000~\euro/kW to 510~\euro/kW to reflect forecasts for 2030 made by the Danish Energy Agency in 2019 \cite{dea2019} (these assumptions are conservative given that some studies saw a cost of 460~\euro/kW for utility solar in Europe in 2019 \cite{doi:10.1002/pip.3189}; with our assumptions the LCOE of wind and solar are still above the average reverse auction results in Germany in 2019).  A table of technology assumptions can be found in \ref{sec:assumptions}.  In order to concentrate on the interaction of market policy and market prices, the wind and solar costs are fixed for the simulations and no learning effects are applied for high penetrations of wind and solar that might reduce costs. In addition, we remove the options for new nuclear and CCS when we focus on the penetration of variable renewable energy under different policies, so that we can achieve the full range of penetrations for the comparison; for the same reason we remove the options for solar, wind and CCS when we focus on nuclear penetration under different policies. Without these removals, each technology would only rise to its particular share in the cost-optimal mix of low carbon technologies for a given CO$_2$ policy; since this mix depends strongly on the model assumptions, and has been well studied in the literature \cite{VERBRUGGEN20084036,HAMACHER2013657,sepulveda2018,kan2020}, we avoid allowing the low-carbon technologies to compete in this study. The removal of the options for new nuclear and CCS for electricity generation can also be seen as representative of the policy environment in countries like Germany with regard to these technologies.

Transmission capacities between countries are fixed at the net transfer capacities (NTC) values from summer 2010. Following \cite{HIRTH2013218} a discount rate of 7\% is applied. To ensure that additional constraints do not distort the theoretical picture developed in the previous section, we do not model unit commitment or assume a baseload premium (whereby nuclear, coal and lignite run even if their variable cost is higher than the market price), nor do we model reserve requirements or revenue from reserve markets. To avoid interference with the CO$_2$ policy we introduce here, we also remove the CO$_2$ price of 20~\euro/tCO$_2$ assumed as a default in \cite{HIRTH2013218}.

In \ref{sec:validation} we compare the results for the relative market value to the results from \cite{HIRTH2013218}, with and without the technology assumption changes. We find good agreement between the models.

To explore the impact of flexibility, in some scenarios we also allow the expansion of the transmission grid and the installation of new storage in the form of batteries and underground hydrogen storage (based on electrolysis of water and hydrogen turbines to feed back into the grid).

\section{Simulation results}
\label{sec:numerical_results}
\begin{figure}[!t]
\centering
    \includegraphics[trim=0 0cm 0 0cm,width=\linewidth,clip=true]{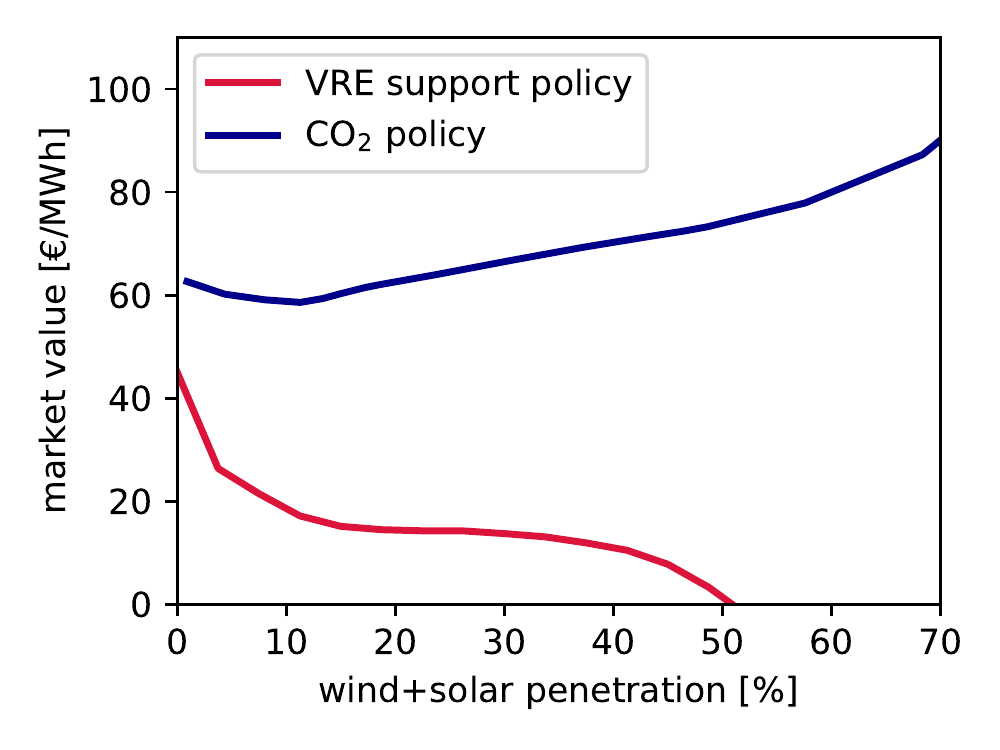}
\caption{Comparison of the market value of wind and solar as their combined penetration is mandated using (i) a VRE support policy and (ii) a CO$_2$ policy.}
\label{fig:comparison}
\end{figure}

\begin{figure}[!t]
\centering
    \includegraphics[trim=0 0cm 0 0cm,width=0.89\linewidth,clip=true]{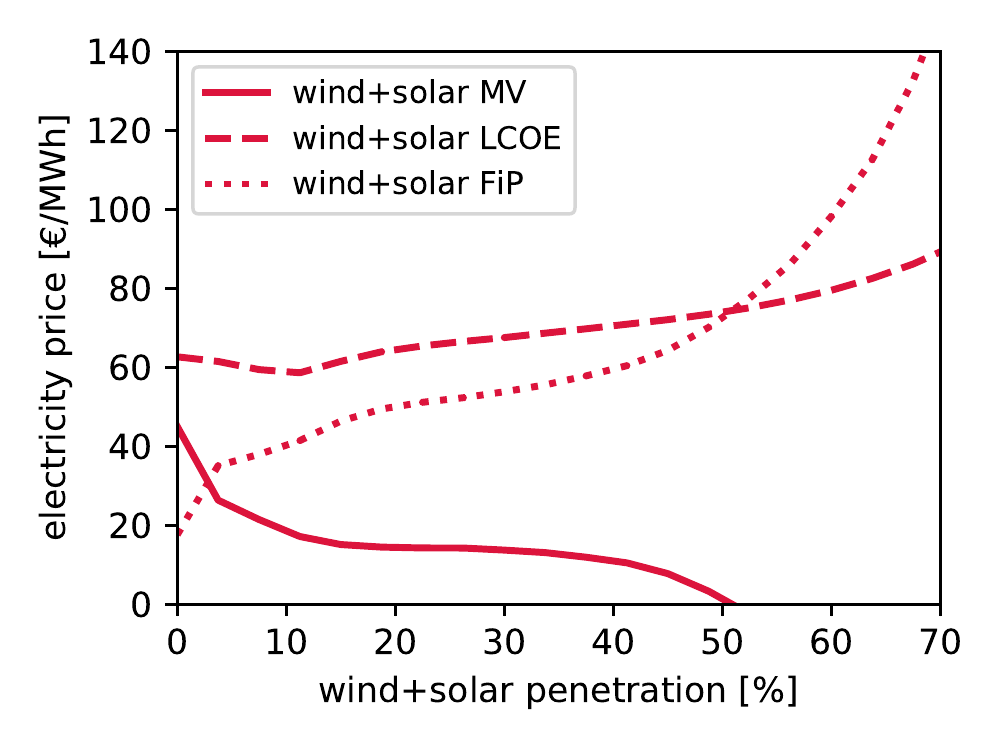}
\caption{Market quantities under a VRE support policy as the penetration of wind and solar energy covering electricity demand is increased. In this case there is no additional flexibility from storage or transmission reinforcement.}
\label{fig:mwh-wind-solar-no_storage}
\end{figure}

\begin{figure}[!t]
\centering
    \includegraphics[trim=0 0cm 0 0cm,width=\linewidth,clip=true]{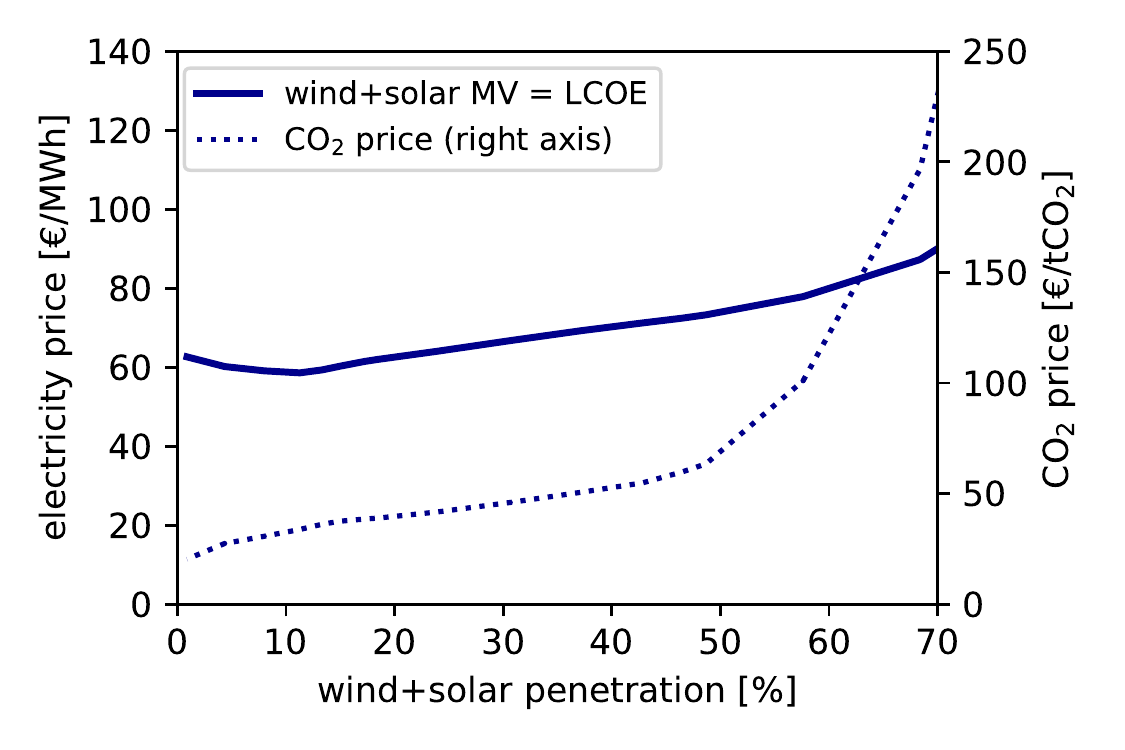}
\caption{Market quantities under a CO$_2$ policy as the average CO$_2$ emission factor is reduced, forcing up the wind and solar penetration.}
\label{fig:mwh-pen-co2-no_storage}
\end{figure}

\subsection{The market value of wind and solar depend on the policy measure}

We contrast two main cases, one where VRE generation is driven by a constraint on minimum penetration level (\emph{VRE support policy}) and another case where VRE generation is driven indirectly by a cap on CO$_2$ emissions (\emph{CO$_2$ policy}), which makes fossil-fuelled generation more costly and thus draws VRE generation into the cost-minimal solution. Technically, the VRE support policy is implemented by a constraint in the optimisation model (equation \eqref{eq:vre}) that mandates a certain share of the demand be fulfilled by wind and solar. The CO$_2$ policy is implemented with a constraint on the maximum CO$_2$ emissions, which corresponds to a tax on CO$_2$-emitting generators (equation \eqref{eq:co2}).

The resulting market value (MV) at penetration levels for wind and solar between 0 and 70\% for these two cases are shown in Figure \ref{fig:comparison}. The results for the VRE support policy case confirm what is widely seen in the literature \cite{borenstein2008,kopp2012,mills2013,HIRTH2013218,gilmore2014,MILLS2015269,gowrisankaran2016,WINKLER2016464}: MV declines with rising penetration, eventually dropping to zero at a VRE penetration of 50\%. The CO$_2$ policy shows a quite different trend: the MV dips slightly, then increases gently up to just over 80 \euro/MWh at 70\% penetration.

This shows clearly that market value behaves differently depending on the policy used to reach a given level of wind and solar generation.

Now we expand upon the results for each policy in detail.

\subsection{Market value with VRE support policy}

Figure \ref{fig:mwh-wind-solar-no_storage} shows the behaviour of the MV, LCOE and Feed-in-Premium (FiP) $\mu_s$\footnote{As outlined in Section \ref{sec:retproof}, the Feed-in-Premium (FiP) $\mu_s$ is the dual, or shadow price, of the VRE constraint \eqref{eq:vre}.} for the VRE support policy.

The LCOE remains approximately constant, dipping first and then rising gently. The dip occurs because of the changing mix of wind and solar, which have different LCOEs: first wind is preferred, which has a higher LCOE but a more regular profile, then solar increases, which has a lower LCOE, before wind takes over again at higher penetrations. The rise in LCOE reflects a preference for wind at higher penetrations, as well as curtailment which lowers the total generation in the denominator of the LCOE.

The FiP has to make up the difference between the MV and LCOE, and thus rises accordingly. The FiP is always positive because the equilibrium solution without the VRE support policy does not contain wind and solar (since the cost of generation from fossil fuels is so low in the model). The MV can reach zero and even become negative, since the FiP can force market prices to be negative in some hours; simulations where negative prices are forbidden are presented in \ref{app:ret-zero}.

\subsection{Market value with CO$_2$ policy}

Figure \ref{fig:mwh-pen-co2-no_storage} shows the MV, LCOE and CO$_2$ tax $\mu_{\textrm{CO}_2}$ for the CO$_2$ policy.\footnote{As outlined in Section \ref{sec:co2proof}, the CO$_2$ tax $\mu_{\textrm{CO}_2}$ is the dual, or shadow price, of the CO$_2$ constraint \eqref{eq:co2}.}

Since wind and solar have no direct CO$_2$ emissions, by equation \eqref{eq:main:co2} the LCOE is exactly equal to the MV. The dip and gentle rise of LCOE has the same explanation as for the VRE support policy (see Appendix Figure \ref{fig:mwh-pen-co2} for the changing shares of wind and solar).

The CO$_2$ price required to induce a given VRE penetration rises to 70~\euro/tCO$_2$ at a penetration of 50\%, before rapidly rising to above 220~\euro/tCO$_2$ at 70\%. 70~\euro/tCO$_2$ is more than the 55~\euro/tCO$_2$ peak seen in early 2021 for CO$_2$ certificate prices in the European Union Emissions Trading System (ETS), but is less than the 129~\euro/tCO$_2$ price expected in 2030 in order to reach the targets of the European Green Deal \cite{Pietzcker2021}, and considerably less than the 195~\euro/tCO$_2$ damages due to climate change estimated in 2020 with a 1\% rate of time preference by the German Environment Agency \cite{uba}. Until 50\% penetration the behaviour of the system under a CO$_2$ policy only requires moderate changes. Beyond 50\%, the lack of additional flexibility options makes CO$_2$ mitigation more expensive.

\begin{figure}[!t]
\centering
    \includegraphics[trim=0 0cm 0 0cm,width=\linewidth,clip=true]{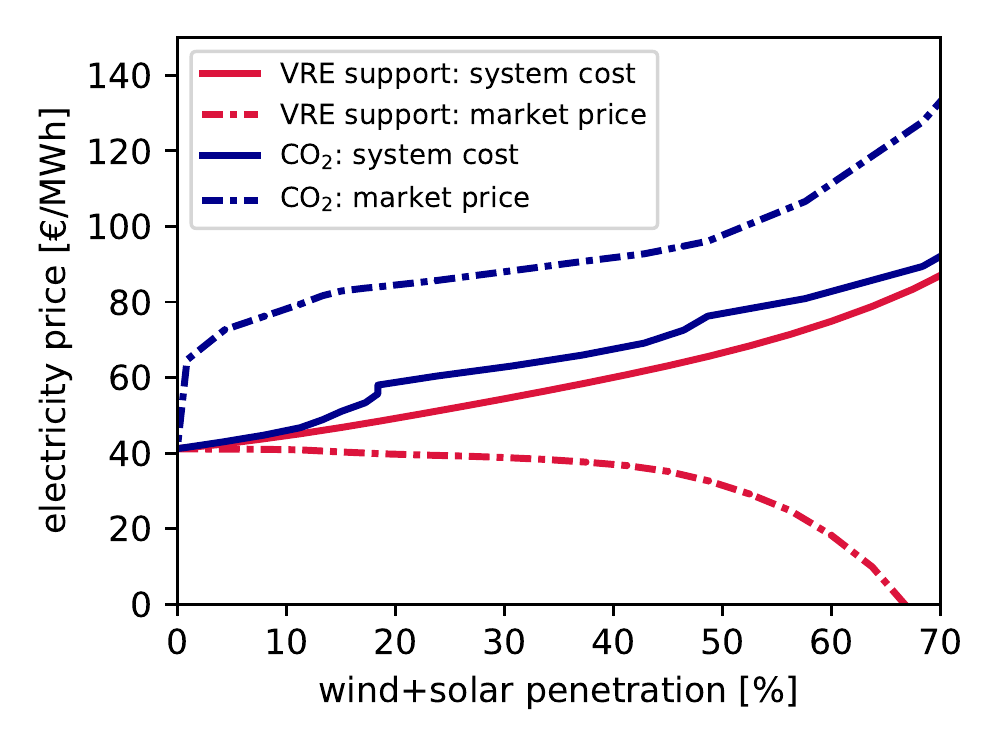}
\caption{Comparison of average system generation cost (excluding CO$_2$ price) and average market price for the VRE support and CO$_2$ policies without flexibility. }
\label{fig:sys_cost-comparison}
\end{figure}

\subsection{System cost and market price}

Figure \ref{fig:sys_cost-comparison} compares the rising average system generation cost for the two policies, including all capital and marginal costs but excluding subsidies and the CO$_2$ price. The costs rise at a similar rate with VRE penetration, implying that both policies achieve similar effects. Costs are slightly higher with a CO$_2$ policy because increasing wind and solar penetration is not the only cost-effective measure to reduce CO$_2$ emissions: switching from lignite and coal to natural gas is prioritised before VRE capacity at some penetrations. These measures make the system more expensive than the VRE support policy case for a given VRE penetration. (If we compare the policies based on CO$_2$ emissions, then the CO$_2$ policy is naturally more efficient at reducing emissions, see Appendix Figure \ref{fig:sys_cost-comparison-co2}.)

Figure \ref{fig:sys_cost-comparison} also shows how the load-weighted average electricity price changes with penetration for each policy. For the VRE support policy, prices are depressed by the merit order effect when VRE generate and by negative prices when VRE are price-setting, as discussed in Section \ref{sec:retproof}. For the CO$_2$ policy the merit order effect of VRE generation is counter-acted by the increasing cost of fossil-fuelled generation, pushing up prices when these generators are price-setting (see Section \ref{sec:co2proof}).\footnote{If the CO$_2$ policy is replaced by a tax on non-VRE generation, the system costs will be identical and the market prices are lifted by a constant factor, as discussed in \ref{sec:lift}. In this case the effect of the CO$_2$ policy is similar because the non-VRE generators all emit CO$_2$, albeit at different rates.}

For the VRE support policy, consumers pay less than the generation cost, since the difference between the average market price and the generation cost is accounted for by the external subsidies paid to VRE generation. For the CO$_2$ policy, consumers pay more than the generation cost, since they must also pay for CO$_2$ emissions according to the prevailing CO$_2$ price. Both the costs of the VRE subsidies and the revenues from the CO$_2$ tax can be passed on to consumers, so that consumers only pay the average system cost in the end, thereby evening out the difference between the policy regimes from the consumers' perspective.\footnote{This is only true for the model setup of perfectly price-inelastic demand; the difference would be more significant with elastic demand.}

\subsection{Including transmission and storage flexibility}

\begin{figure}[!t]
\centering
    \includegraphics[trim=0 0cm 0 0cm,width=\linewidth,clip=true]{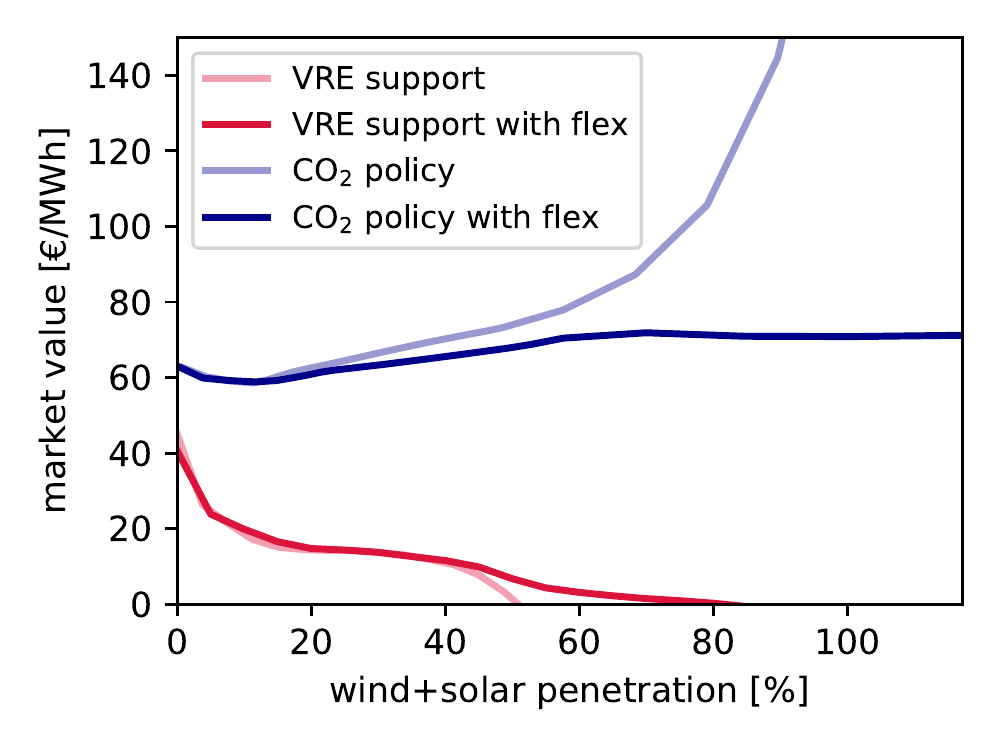}
\caption{Market value under VRE support and CO$_2$ policies as wind and solar penetrations rise, for scenarios with and without additional flexibility from transmission and storage. The penetration of VRE as a fraction of demand goes beyond 100\% to 117\% because VRE must also cover storage losses.}
\label{fig:co2-compare}
\end{figure}

\begin{figure}[!t]
\centering
    \includegraphics[trim=0 0cm 0 0cm,width=\linewidth,clip=true]{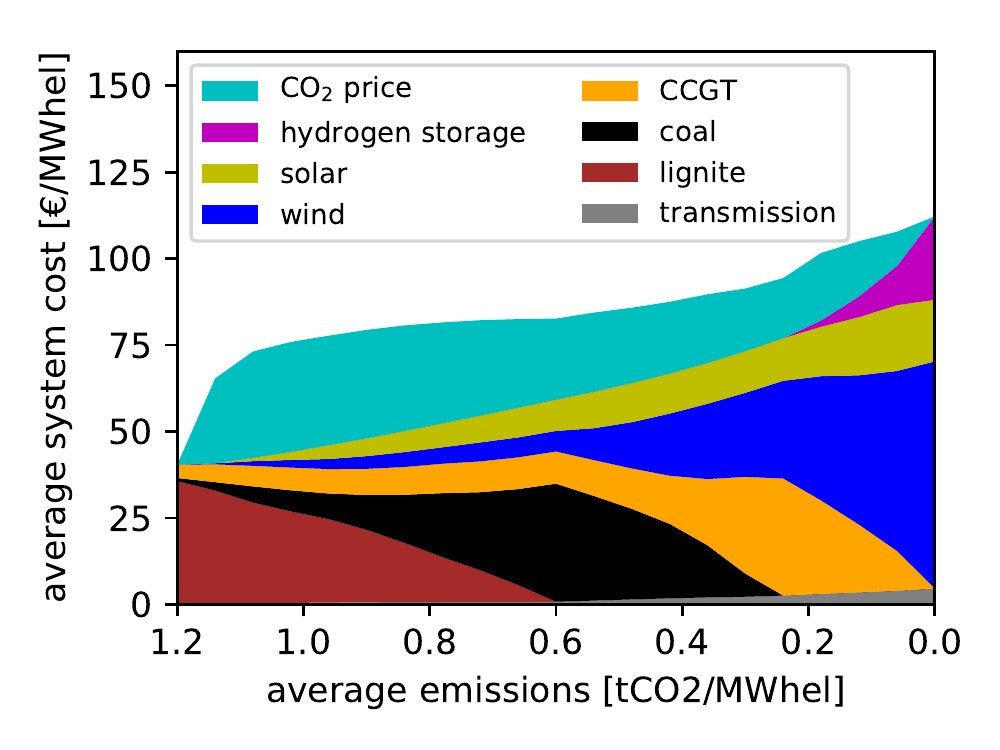}
\caption{Breakdown of average system cost (total cost divided by total load) under a CO$_2$ policy as the CO$_2$ budget is tightened, for a scenario with transmission expansion as well as short- and long-term storage. The average system cost, including the CO$_2$ price, is equal to the load-weighted average market price by equation \eqref{sec:total_costs}.
}
\label{fig:sys_cost}
\end{figure}

If additional flexibility options are made available to the investment optimisation, the market value under a CO$_2$ policy remains regular all the way up to full VRE penetration. Flexibility in this case includes the option to build new transmission capacity between the countries, as well as the availability of both battery storage and hydrogen storage. The results for MV of VRE with and without flexibility are shown in Figure \ref{fig:co2-compare}. Without flexibility, the MV increases strongly above 70\% because of high curtailment that depresses the LCOE. High curtailment reflects the mismatch between VRE and demand profiles. With flexibility, the MV rises slowly before plateauing at around 71~\euro/MWh. When VRE covers all of the demand and storage losses, the average total system cost is higher at 114~\euro/MWh, reflecting the cost of additional flexibility options, in this case primarily the hydrogen storage. The CO$_2$ price rises from 55~\euro/tCO$_2$ at 50\% to 165~\euro/tCO$_2$ at 100\% penetration (less than the climate damages of 195~\euro/tCO$_2$ estimated in 2020 with a 1\% rate of time preference in \cite{uba}).

The breakdown of system cost by component in Figure \ref{fig:sys_cost} shows the substitution of technologies as the CO$_2$ limit is tightened.
Hydrogen storage is critical for removing the final emissions from the system, since hydrogen stored underground can provide power when wind and solar feed-in is low for multiple days. The fact that wind and solar dominate system costs at the same time as dominating energy generation guarantees a relative market value (RMV) close to unity according to equation \eqref{eq:rmvid}. Even at full VRE penetration, the RMV of wind and solar only drops to 0.62, see Appendix Figure \ref{fig:comparison-rmv}. Similar results have also been shown in models coupled to building heating and transport, where demand response from electric vehicles and heat pumps, as well as cheap storage of heat, hydrogen and methane, help to support prices using price arbitrage and keep the RMV close to unity \cite{Brown2019,haertel2021,bernath2021,boettger2021,Ruhnau2020Market}. Flexible demand and storage bid up prices by providing extra demand when VRE are abundant, and lower prices by reducing demand when VRE are scarce.

It is sometimes assumed that prices become singular in a system based entirely on wind and solar, alternating between zero during VRE abundance and very high levels during VRE scarcity. In \ref{app:duration} we show that this is not the case by looking at the price duration curves in the system. As fossil-fuelled generation is pushed out of the system, storage and transmission arbitrage start to set the prices, removing singular prices with demand bids when VRE is abundant and supply bids when VRE is scarce. The distribution of hours per year in which wind and solar recoup their costs barely changes as the CO$_2$ budget is lowered.

\subsection{Support versus CO$_2$ policies for nuclear power}

\begin{figure}[!t]
\centering
    \includegraphics[trim=0 0cm 0 0cm,width=\linewidth,clip=true]{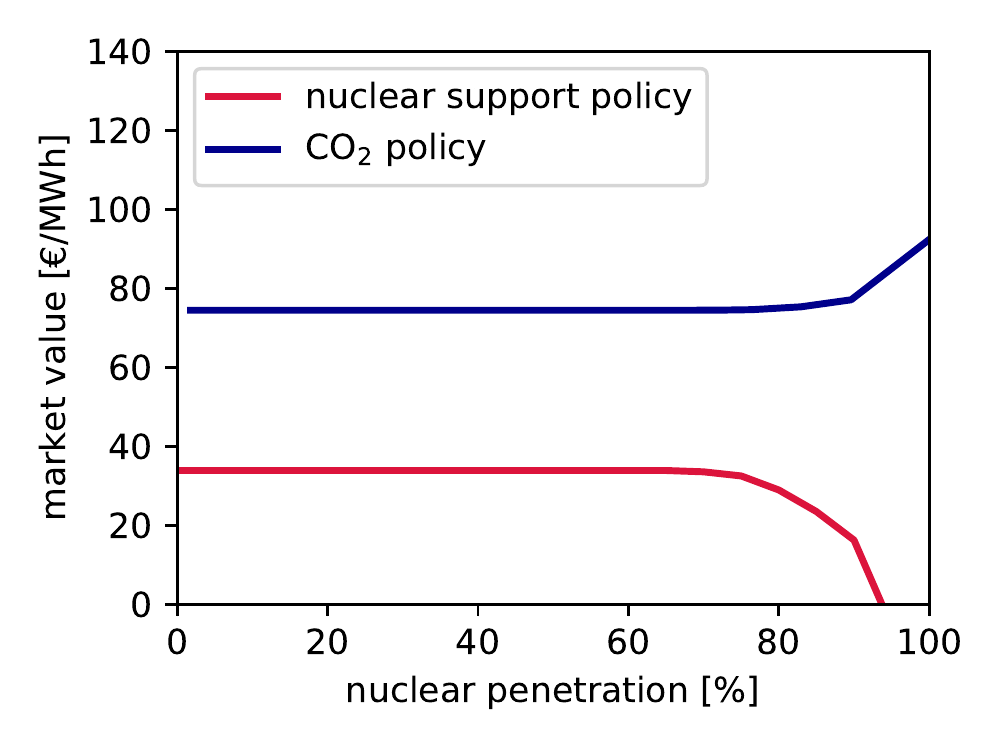}
\caption{Comparison of the market value of nuclear as its penetration is mandated using (i) a nuclear support policy and (ii) a CO$_2$ policy.}
\label{fig:comparison-nuclear}
\end{figure}

While we have focused on wind and solar generation, reflecting the focus of the literature on market value, exactly the same considerations apply in the case of non-variable low-emissions technologies like nuclear.

Figure \ref{fig:comparison-nuclear} shows the behaviour of nuclear's market value under a nuclear support policy and under a CO$_2$ policy. As described in Section \ref{sec:numericalmodel}, we have removed the options to build wind, solar and CCS so that we can focus on nuclear penetration versus fossil-fuelled generation. It is assumed that nuclear can be built to operate flexibly \cite{Jenkins2018}.

With a technology-specific support policy pushing a fixed share of nuclear power, the market value of 34~\euro/MWh at low penetrations is much lower than the LCOE of 74~\euro/MWh, implying that it requires a subsidy of 40~\euro/MWh to compete with the other baseload power source in the system, lignite. The support policy has pushed nuclear's market value below the point of cost recovery, causing it to cannibalise its own revenue. At higher penetrations the market value drops all the way to zero, since the cost characteristics of nuclear are not suitable to match the full variability of the demand. For nuclear it is the variability of demand, rather than of supply in the case of VRE, that combines with the support policy to push the market value to zero due to the mismatch of the supply and demand profiles.

With a CO$_2$ policy drawing nuclear into the system, the market value matches the LCOE, so that nuclear exactly makes back its costs from the market. The LCOE rises at higher penetrations as the capacity factor of nuclear drops to match the variable demand. Initially a CO$_2$ price of 34~\euro/tCO$_2$ is necessary for nuclear to displace lignite and reach a share of 68\% of electricity generation. To reach 90\% penetration, the CO$_2$ price must rise to 69~\euro/tCO$_2$, and even higher for higher penetrations (see Appendix Figure \ref{fig:mwh-pen-nuclear}).

Both the cannibalisation effects and the drop in market value with penetration under a support policy, as well as the fundamentally different behaviour under support versus CO$_2$ policy regimes, follow the same pattern as the case of VRE technologies (compare Figures \ref{fig:co2-compare} and \ref{fig:comparison-nuclear}). This was to be expected, given that the theoretical considerations in previous sections are technology-neutral. The decline of market value under a support policy happens for all technologies regardless of their variability or other techno-economic characteristics, and is thus primarily a policy-dependent phenomenon. The only difference between dispatchable and variable technologies is that this decline happens faster for VRE because they do not match the shape of the variable demand as well as dispatchable technologies, making the impact of variability on market value only a secondary effect to policy, since it affects the rate of change rather than the direction.

\section{Discussion}
\subsection{The mechanisms underlying MV decline}

This paper contrasts the impact of support and CO$_2$ policies on the market value (MV) of wind and solar. We find that the MV of wind and solar decreases strongly under VRE support policies, whereas under CO$_2$ policies the MV remains high enough for wind and solar to cover their costs from the market. Thus, the declining MV of VRE that has been observed in previous literature \cite{kopp2012,HIRTH2013218,MILLS2015269,WINKLER2016464,BLAZQUEZ20181,Hirth2016} is caused primarily by the implicit assumption of policy regime, and not, as has been claimed before \cite{HIRTH2013218, BLAZQUEZ20181}, by the variability of wind and solar, although the variability can contribute towards the speed of decline.

Many of the papers on market value in long-term equilibrium models achieve rising shares of wind and solar either by exogenously fixing the capacity of wind and solar \cite{borenstein2008,kopp2012,mills2013,gilmore2014,MILLS2015269,gowrisankaran2016} or by forcing them into the system with a constraint \cite{HIRTH2013218,hirth2014,pahle2015,WINKLER2016464,HIRTH2016210,hirth2016b}\footnote{Note that Lamont \cite{LAMONT20081208} does not follow this approach, but draws in wind and solar by reducing their capital costs, thereby simulating the effects of technological learning and guaranteeing that costs are covered from the market.}, without always making clear that forcing a share of wind and solar is equivalent to a VRE support policy.
%The equivalent feed-in premium (equal to the shadow price of the constraint) distorts the equilibrium and forces down the market value of wind and solar.
For example, Hirth \cite{HIRTH2013218} claims to `identify and quantify the impact of prices and policies on the market value of VRE', but the policies do not include FiTs or renewable portfolio standards. Such policies are, however, implicit in the study design since it increases the penetration level beyond the equilibrium share, which would never come about without technology-specific policies.\footnote{
Consider what would happen in the long-term equilibrium without any policy intervention: there would be no market value decline, since the technology would settle at its optimal long-term share with MV $=$ LCOE, a share based on its variability, costs and the rest of the generation mix. It takes policy to distort this equilibrium, thereby altering the share and the market values.
} Therefore, their conclusions that `the market value of both wind and solar power is significantly reduced by increasing the market shares of the respective technology' is not universally true, but only under the assumption that the market shares are increased using a technology-specific support mechanism. The paper by Winkler et al. \cite{WINKLER2016464} has a similar set-up, where the increase in penetration level, i.e. the implicit technology support policy, is part of the study design, yet not recognised as a policy.
%Winkler et al.\cite{winkler2016impact}, for example, identify a $CO_2$ price as the most powerful driver to prevent MV from plummeting. Both Hirth \cite{HIRTH2013218} and Zipp \cite{zipp2017marketability} list a higher emission price signal as one of the ways in which the decreasing market value may be mitigated, but do not dwell on the mechanism by which this happens. Hogan \cite{HOGAN201755} notes that one of the roots for the missing money problem is the failure to understand the consequences of policies for support of zero-carbon generation technologies, but his analysis is limited to the mechanisms to ensure reliability.

Much of the literature has viewed a CO$_2$ price as only one among many other mechanisms by which the MV decline can be mitigated under a VRE support policy regime \cite{kopp2012,mills2013,HIRTH2013218,gilmore2014,hirth2014,gowrisankaran2016,WINKLER2016464,ZIPP20171111}.
Since these papers leave the CO$_2$ price fixed, they still observe a declining market value as the subsidy for VRE is increased, even with a high CO$_2$ price of 100~\euro/tCO$_2$ and the removal of nuclear and CCS \cite{HIRTH2013218,hirth2014}.
In contrast, we replace the subsidy with a rising CO$_2$ price as the primary mechanism to draw in wind and solar, thus guaranteeing that there is no decline in market value at all.

A policy regime of subsidy for a particular technology will drive down its market value regardless of its variability, location, fuel cost or the rest of the generation mix. Studies that demonstrate declining market value are simply reproducing this basic point, except that they obscure the nature of the subsidy by including it only implicitly as a constraint that forces in the technology share. Our contribution is to explain the mechanism by which this constraint acts as a subsidy to depress prices, showing it both in theory and in simulations. In addition, we also show that there is nothing particular about variable generators with respect to market decline with increased forced share. This also happens to nuclear, see Figure \ref{fig:comparison-nuclear}. It is the fact that a technology is forced beyond its equilibrium share that causes the decline: the particular techno-economic properties of the technology (variability, ratio of CAPEX to OPEX) and the rest of the mix (the other available technologies and their properties) only affect the speed of decline. A CO$_2$ policy that draws in wind and solar with a rising CO2 tax causes no decline, see Figure \ref{fig:mwh-pen-co2-no_storage}. This is true for wind, solar and other low-CO$_2$ technologies like nuclear, see Figure \ref{fig:comparison-nuclear}.\footnote{Remember from Section \ref{sec:numericalmodel} that in the CO$_2$ policy simulations, other low-carbon technologies were removed to achieve the full range of penetrations. With competing technologies, cost-recovery from the market under a CO$_2$ policy still applies for a mix of technologies following the theory in Section \ref{sec:math_theory}.} The variability is only a secondary characteristic that affects the  \emph{rate} of decline of MV under support policies, which is faster for VRE than for nuclear. We thereby demonstrate that the policy mix is the primary mechanism affecting market value (since it affects the fundamental direction of market value change), while variability, fuel cost and generation mix only affect the rate of change of market value in the policy regime.

We grant a whole host of strategies, which can be collectively labeled `flexibility measures'\footnote{These have been individually investigated in previous literature and may be DSM measures \cite{pahle2015, WINKLER2016464}, storage \cite{HIRTH2013218}, hydro power \cite{HIRTH2016210,Obersteiner2010, Tveten2016} or transmission extensions \cite{HIRTH2013218, Obersteiner2010}.}, as having the potential to \emph{dampen} the MV decline under a VRE support policy, yet not solving the bigger issue of the `cannibalisation effect' \cite{HOGAN201755}, which applies as much for VRE as for nuclear if they are subsidised.

%Policy driven by a CO$_2$ price, however, has the potential to prevent any MV decline, in addition to its other advantages of being technologically neutral and economically efficient \cite{RePEc:aen:journl:2006v27-03-a03, FELDER2002107}.
\subsection{Policy implications}\label{sec:policy}

The main policy implication is that policy makers should not see market value decline under VRE support policies as an indication that variable renewable energy is hitting fundamental integration limits. Thus we oppose the notion expressed e.g. in Hirth and Radebach \cite{Hirth2016}, where they claim that: `variable renewables face a substantial difficulty in becoming economical at high market shares. Without fundamental technological breakthroughs, a deep decarbonisation of power systems will be hard to achieve based on wind and solar power alone. Other supplementary low carbon technologies are likely to be needed.'

MV decline is a result of policy choices rather than an intrinsic property of VRE. In particular, it is a result of choosing not to value the low CO$_2$ emissions of wind and solar inside the market, but to subsidise VRE outside of the market. The strong dependence of market value on the policy regime (and on the rest of the system composition) means that market value should be used with a keen awareness of its limitations. Just as the LCOE metric does not provide a complete picture of the cost performance of technologies \cite{competitiveness2020}, market value should be used in concert with other metrics when comparing technologies, such as the effect on total system cost.

This paper focuses on the effects of policy on market value rather than on the desirability of the policy measures themselves.\footnote{For such an investigation, more sophisticated methods, which account for a broader technology selection, demand elasticity, inter-temporal dynamics, learning effects, path dependencies \cite{grubb2021}, risk averse agents and political economy aspects would be necessary.} While a CO$_2$ policy may be more efficient to reduce CO$_2$ emissions in theory, there are many situations where VRE support policies are preferable to CO$_2$ policies, such as when encouraging research, development and deployment to lower costs through learning, reducing investor risk, or because in some regions subsidies enjoy more political support than taxes. In these cases the measure of successful integration should be the total cost of the system rather than market value, since the system cost can be calculated regardless of the market structure. Comparison to the economically efficient solution with a CO$_2$ policy may also provide useful guidance.\footnote{In the model setup in this paper the system costs for each policy for a given level of CO$_2$ emissions are quite close, see Appendix Figure \ref{fig:sys_cost-comparison-co2}.}

%\subsection{Effects on investors}

In particular the impact on financial risk, and thus investor behaviour, may differ substantially between support and CO$_2$ policies: CO$_2$ policies send a market signal to encourage low-emission generation, whereas one of the main purposes of support policies is to provide investor certainty for capital-intensive investments that might otherwise be subject to market risks from fluctuating electricity and CO$_2$ prices \cite{HIROUX20103135,Held2019}. Lower risk means lower financing costs, which feeds through to a lower LCOE and a lower system cost \cite{Schmidt2019, Egli2019, BUTLER20081854}. A hybrid policy framework can provide both of these benefits: a CO$_2$ price to support low-emission generation, and a guaranteed per-kWh feed-in premium for VRE generation. If the CO$_2$ price is sufficiently high, then the feed-in premium for a given share of VRE,  $\mu_\Gamma$ in equation \eqref{eq:main}, can be close to zero. This provides minimal market distortion, costs consumers very little, but reduces investor risk and thus lowers financing costs.

\subsection{Negative prices}

Under the VRE support policy, electricity prices may become negative because it is rational for VRE generators to offer negative bids, since they are subsidised for their feed-in regardless of the market price. VRE generators have an effective bid of their running cost minus a feed-in premium equivalent to the shadow cost of the VRE-constraint, as discussed in Section \ref{sec:retproof}.\footnote{Note that this cause of negative prices is distinct from other causes, such as unit commitment or network constraints.} A similar construct was
used and negative prices were observed in \cite{pahle2015,gl2015}. In setups where the available energy rather than the dispatched energy is
constrained \cite{HIRTH2013218} VRE support does not cause negative prices because the constraint is equivalent to subsidising the capacity rather than the energy generation, see \ref{sec:retcapproof}.

In reality, some countries have policies that withdraw subsidies when prices go negative for a sufficient time (4 hours in the case of  Germany for generators built from 2021
\cite{EEG2021}). Under such policy regimes, it would be rational for the producer never to bid in less than its running costs to the market, and thus the
market prices would be higher. Results for support policies where negative prices are forbidden are provided in \ref{app:ret-zero}. Under the VRE support policy the market values still decline well below the cost recovery point with increasing VRE penetration, but do so more slowly and do not turn negative.

   % \item \emph{The model used here cannot account for the inertia to reshape the energy system}. Should VRE policies dominate the incentives for CO$_2$ mitigation also in the future, it is likely that both the renewable and the thermal power mix be highly suboptimal from the point of view of CO$_2$ reduction. The reason is that VRE policies do not incentivize neither other CO$_2$ neutral generation (such as biomass, CCS and nuclear), nor does it foster a sound capacity mix in the rest of the system. In fact, coal is part of the optimal capacity mix even at 50\% or more VRE in the power mix (Figure \ref{fig:sys_cost}). In the real system, such a suboptimal capacity mix has a long lifetime, creating an unnecessarily costly transition.

%Taken together, these three mechanisms illustrate that, in absence of a CO$_2$ price level with complete certainty, the policy mix appropriate to help decarbonize the power sector is not obvious.
\subsection{MV under different policies in reality: the example of Germany}
\begin{figure}[!t]
\centering
    \includegraphics[trim=0 0cm 0 0cm,width=\linewidth,clip=true]{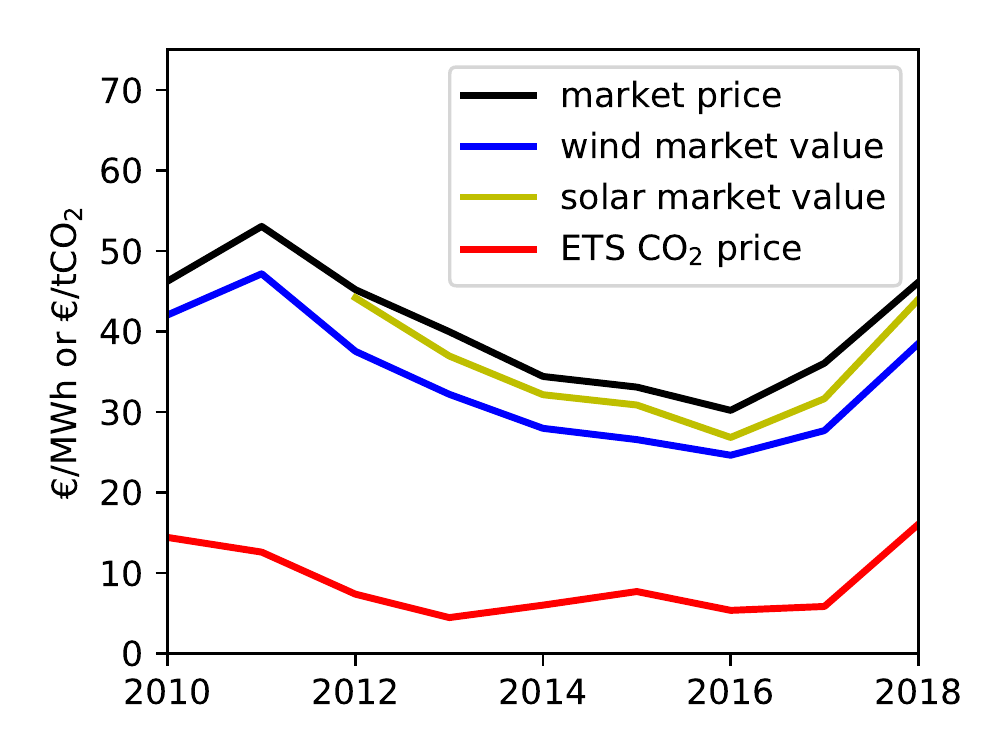}
\caption{Market data from Germany, 2010-2018. As the CO$_2$ price rose towards the end of the period, so have the average market price and market values of wind and solar. Data from \cite{OPSD}.}
\label{fig:real-data}
\end{figure}

It is undisputed that the revenues from sales on the market for VRE generators have decreased in the real world, which has been shown in several studies on historical data for electricity prices
\cite{Sensfuss2007_1000007777,SENSFU20083086,7080941,Figueiredo2017, ozdemir2017integration,Hirth2018143,LopezProl2020,MILLS2015269,HIRTH2013218}.
While this is partly due to short-term effects as the rest of the system takes time to reach a new equilibrium with VRE, it also reflects policy choices. In Germany, subsidies were used to increase the share of VRE in electricity to around 26\% in 2018. The average power price and market values of wind and solar fell from 2011 to 2016, as can be seen in Figure \ref{fig:real-data}. This is an effect of the FiT policy regime (which is a VRE support policy) combined with falling fuel prices. However, from 2016 onwards, prices increase again, with market values approaching current German LCOEs for wind and solar. This has been attributed to an increasing CO$_2$ price on the EU Emissions Trading System (ETS) market (a CO$_2$ policy) \cite{Agora-JAW-2019}. Thus the German prices (Figure \ref{fig:real-data}) may be interpreted as a supporting argument for our evaluation of the effect when VRE support policies dominate (up until 2016) and when CO$_2$ policies have a strong effect (after 2016).

\subsection{Limitations of this study}

In this study we have focused on the mechanisms connecting cost, price and policy in the long-term. The theory holds regardless of any specific technology assumptions, but many of the simulation results depend on the background system choices we have made. The base model excludes new nuclear, CCS, price-elastic demand, demand-side management (DSM), coupling to transport or heating, hydroelectric dams and interconnection across the whole of Europe. A larger geographic scope or the inclusion of hydroelectric dams, DSM or sector coupling would expand the flexibility mechanisms and thereby dampen the decrease of MV under VRE support policies and decrease system cost \cite{Brown2018,Brown2019}. New nuclear or CCS could compete with VRE under a CO$_2$ policy and limit the ultimate VRE penetration, depending on the costs, but they would not affect the conclusions on market value. The approach of grouping generators into representative classes for each technology, which we took over from \cite{HIRTH2013218}, is standard practice in long-term equilibrium modelling but leads to a strongly simplified step-wise merit order curve compared to the smoother curve that would arise from a wide variety of generator types. This does not affect our theoretical results for the long-term equilibrium, but may impact the rate of market value decline in a more realistic short-term model.

\section{Conclusions}

The market value of wind and solar (VRE) depends strongly on the policies used to promote them. Previous studies have implicitly assumed that direct subsidies are used to force VRE penetration, which have the effect of depressing both their market value and overall market prices. If instead a CO$_2$ price is used to draw in low-emission generation, market values of generators in long-term equilibria are guaranteed to cover the generators' costs. Market values remain stable even at VRE penetrations approaching 100\%, as long as sufficient flexibility from transmission and storage is available in the system.

This means that declining market value under support policies, such as Feed-in-Tariffs or Renewable Portfolio Standards, does not necessarily indicate problems with the market integration of VRE. Declining market value is rather a side-effect of choosing a technology support policy, rather than creating value in the market for technologies with low CO$_2$ emissions. A better measure of market integration is the total system cost, since it can be calculated regardless of the market structure.

By showing the strong dependence of market value on policy choice, we have thus resolved the apparent contradiction between the literature showing market value decline with penetration under support policies, and the literature showing that high penetrations of VRE can be cost-effective under CO$_2$ policies.

\section*{Acknowledgements}

We thank Emil Dimanchev, Tommi Ekholm, Sabine Fuss, Jessica Jewell, Jonathan Koomey, Wolf-Peter Schill, Richard Schmalensee, Johannes Schmidt, Afzal Siddiqui, Patrik Söderholm, Thomas Sterner and Alexander Zerrahn for helpful discussions,
suggestions and comments.  T.B. acknowledges funding from the
Helmholtz Association under grant no.~VH-NG-1352. The responsibility for the contents lies with the
authors.

\appendix

\section{Karush-Kuhn-Tucker (KKT) conditions}\label{sec:kkt}

In this section we set the signs and notation for the Karush-Kuhn-Tucker (KKT) conditions.

We have an objective function over variables labelled by $l$:
\begin{equation}
  \max_{x_l} f(x_l)
\end{equation}
subject to equality ($i$) and inequality ($j$) constraints:
\begin{align}
  g_i(x_l) =  0 \hspace{1cm}\perp \hspace{1cm} \l_i \\
      h_j(x_l) \leq  0 \hspace{1cm}\perp \hspace{1cm} \m_j
\end{align}
We build the KKT Lagrangian:
\begin{equation}
  \cL(x_l,\l_i,\m_j) = f(x_l) - \sum_i \l_i g_i(x_l) - \sum_j \m_j h_j(x_l)
\end{equation}
The KKT conditions are equations satisfied by $x_l$, $\l_i$ and $\m_j$ at the optimum point.

First we have stationarity:
\begin{equation}
 0 =     \frac{\d \cL}{\d x_l} =   \frac{\d f}{\d x_l} - \sum_i \l_i \frac{\d g_i}{\d x_l}  - \sum_j \m_j \frac{\d h_j}{\d x_l}
\end{equation}
then primal feasibility:
\begin{align}
  g_i(x_l) & = 0 \\
    h_j(x_l) & \leq 0
\end{align}
then dual feasibility:
\begin{align}
  \m_j \geq 0
\end{align}
and finally complementary slackness:
\begin{align}
  \m_jh_j(x_l)= 0
\end{align}
(i.e. either $\m_j = 0$ or the inequality constraint is binding $h_j(x_l) = 0$).

\section{Further Proofs}\label{sec:furtherproofs}

\subsection{Single node long-term equilibrium with VRE support policy for available power}\label{sec:retcapproof}

If we add a constraint for a subset $S$ of generators based on the available power before curtailment, as is done in \cite{HIRTH2013218}, rather than the actual dispatched power
\begin{equation}
  -\sum_{s\in S, t} \bar{g}_{s,t} G_s \leq -\Theta  \hspace{0.34cm}\perp \hspace{0.34cm} \m_{\Theta}
\end{equation}
($\bar{g}_{s,t}$ is the hourly capacity factor for generator $s$ at time $t$) then this alters the stationarity equation \eqref{eq:statcapvanilla} for $G_s$ to
\begin{align}
 \frac{\d \cL}{\d G_{s}} = 0 &\Rightarrow  -c_s + \sum_t \bar{g}_{s,t}\bar{\m}_{s,t} +\sum_t  \bar{g}_{s,t} \m_{\Theta} = 0
\end{align}
so that now for renewable generators
\begin{align}
  c_s G_s + \sum_{t} o_{s} g_{s,t}
  & = \sum_{t} \l_t g_{s,t} + \m_{\Theta} G_s \sum_{t} \bar{g}_{s,t} \label{eq:availcosts}
\end{align}
If there is no curtailment, $ \bar{g}_{s,t} G_s = g_{s,t}$ and this becomes the same expression as \eqref{eq:noprofit-vre}.

Since $\m_{\Theta}$ multiplies the capacity $G_s$, this can be interpreted as a subsidy for capacity. (In Section \ref{sec:retproof} we fixed the share of dispatched generation instead, so there it was a subsidy on dispatch.) This means the effective marginal cost is not affected and does not go negative. This makes sense because if capacity is subsidised, generators have no incentive to feed in when prices are negative. Instead, they curtail the available energy.

Dividing \eqref{eq:availcosts} by the total generation $\sum_t g_{s,t}$ we find for $s \in S$
\begin{equation}
  LCOE_s = MV_s  + \m_\Theta \frac{G_s \sum_t\bar{g}_{s,t} }{\sum_t g_{s,t}} \hspace{0.34cm} \forall s \in S
\end{equation}
For $s \notin S$ we have the regular no-profit rule
\begin{equation}
  LCOE_s = MV_s  \hspace{0.34cm} \forall s \notin S
\end{equation}

\subsection{Single node long-term equilibrium with limited installation potentials}

If there are limits on installable potentials for generators
\begin{equation}
  G_{s} \leq G_{s}^{max} \perp \m_{s}^{max}
\end{equation}
then we get
\begin{align}
    \frac{\d \cL}{\d G_{s}} = 0 &\Rightarrow  c_s - \sum_t \bar{\m}_{s,t} + \m_{s}^{max}  = 0
\end{align}
and now
\begin{align}
  c_s G_s + \sum_{t} o_{s} g_{s,t}
  & = \sum_{t} \l_t g_{s,t} - \sum_{s} \m_s^{max} G_s
\end{align}
We have effectively added to the capital cost $c_s$ a cost related to the scarcity of the potential for $G_s$, which drives up the cost. Because the resource is scarce, generators can claim extra revenue for this scarcity, i.e. because there is no alternative, there is extra profit to be obtained from the market.

\subsection{Single node long-term equilibrium with storage}\label{sec:storage}

Suppose we add storage units $r$ with discharging dispatch $g^{\textrm{dis}}_{r,t}$ and power capacity $G^{\textrm{dis}}_{r}$, storing power $g^{\textrm{sto}}_{r,t}$ and capacity $G^{\textrm{sto}}_{r}$, and state of charge $g^{\textrm{ene}}_{r,t}$ and energy capacity $G^{\textrm{ene}}_{r}$. The efficiency from hour to hour is $\eta^{\textrm{ene}}_r$ (for losses due to self-discharge), the storing efficiency is $\eta^{\textrm{sto}}_r$ and the dispatch efficiency is $\eta^{\textrm{dis}}_r$.

We add to the objective function an additional cost term:
\begin{equation*}
  -\sum_{r,\circ} c^\circ_r G^\circ_r =  -\sum_r c^{\textrm{ene}}_r G^{\textrm{ene}}_r -\sum_r c^{\textrm{sto}}_r G^{\textrm{sto}}_r -\sum_r c^{\textrm{dis}}_r G^{\textrm{dis}}_r
\end{equation*}
where the symbol $\circ$ runs over $\{\textrm{ene},\textrm{sto},\textrm{dis}\}$. We assume no marginal costs for the dispatch.

The demand balancing equation \eqref{eq:balance} is modified to:
\begin{equation}
   \sum_a d_{a,t} - \sum_s g_{s,t} - \sum_r g^{\textrm{dis}}_{r,t}  + \sum_r g^{\textrm{sto}}_{r,t}  =  0 \hspace{0.34cm}\perp \hspace{0.34cm} \l_t \hspace{0.34cm} \forall t
\end{equation}
The standard capacity constraints apply:
\begin{align}
    -g^\circ_{r,t}\leq 0 \hspace{1cm}\perp \hspace{1cm} \ubar{\mu}^\circ_{r,t} \hspace{1cm} \forall r,t  \label{eq:storlower}\\
    g^\circ_{r,t} - G^\circ_r \leq 0 \hspace{1cm}\perp \hspace{1cm} \bar{\mu}^\circ_{r,t} \hspace{1cm} \forall r,t \label{eq:storupper}
\end{align}
In addition we have the constraint for the consistency of the state of charge between hours according to how much was dispatched or stored:
\begin{equation}
    g^{\textrm{ene}}_{r,t}- \eta^{\textrm{ene}}_r g^{\textrm{ene}}_{r,t-1} - \eta^{\textrm{sto}}_r g^{\textrm{sto}}_{r,t} + (\eta^{\textrm{dis}}_r)^{-1} g^{\textrm{dis}}_{r,t}  =  0 \hspace{0.14cm}\perp \hspace{0.14cm} \l^{\textrm{ene}}_{r,t} \hspace{0.14cm} \forall r,t  \label{eq:storsoc}
\end{equation}
We assume that the state of charge is cyclic $g^{\textrm{ene}}_{r,-1} = g^{\textrm{ene}}_{r,T-1}$.

From KKT stationarity we get:
\begin{align}
\frac{\d \cL}{\d G^\circ_{r}} = 0 &\Rightarrow - c^\circ_r + \sum_t \bar{\m}^\circ_{r,t}  = 0 \\
    \frac{\d \cL}{\d g^{\textrm{dis}}_{r,t}} = 0 &\Rightarrow  \l_t + \ubar{\m}^{\textrm{dis}}_{r,t} - \bar{\m}^{\textrm{dis}}_{r,t} - (\eta^{\textrm{dis}}_r)^{-1} \l^{\textrm{ene}}_{r,t}  = 0 \\
    \frac{\d \cL}{\d g^{\textrm{sto}}_{r,t}} = 0 &\Rightarrow  -\l_t + \ubar{\m}^{\textrm{sto}}_{r,t} - \bar{\m}^{\textrm{sto}}_{r,t} + \eta^{\textrm{sto}}_r \l^{\textrm{ene}}_{r,t}  = 0 \\
    \frac{\d \cL}{\d g^{\textrm{ene}}_{r,t}} = 0 &\Rightarrow   \ubar{\m}^{\textrm{ene}}_{r,t} - \bar{\m}^{\textrm{ene}}_{r,t} -  \l^{\textrm{ene}}_{r,t} + \eta^{\textrm{ene}}_r \l^{\textrm{ene}}_{r,t+1}   = 0
\end{align}

The zero-profit rule for storage proceeds the usual way:
\begin{align}
  \sum_\circ c^\circ_r G^\circ_r & =  \sum_{\circ,t} G^\circ_r\bar{\m}^\circ_{r,t}  =   \sum_{\circ,t} g^\circ_{r,t}\bar{\m}^\circ_{r,t} \nonumber \\
   = & \sum_t \left[ \l_t g^{\textrm{dis}}_{r,t} -(\eta^{\textrm{dis}}_r)^{-1} \l^{\textrm{ene}}_{r,t}  g^{\textrm{dis}}_{r,t}
  -\l_t g^{\textrm{sto}}_{r,t} + \eta^{\textrm{sto}}_r \l^{\textrm{ene}}_{r,t} g^{\textrm{sto}}_{r,t} \right.\nonumber \\
  & \left. \hspace{.5cm}-\l^{\textrm{ene}}_{r,t}g^{\textrm{ene}}_{r,t} + \eta^{\textrm{ene}}_r \l^{\textrm{ene}}_{r,t+1}g^{\textrm{ene}}_{r,t} \right] \nonumber \\
   = & \sum_t \l_t \left[ g^{\textrm{dis}}_{r,t} - g^{\textrm{sto}}_{r,t}  \right] \nonumber \\
   & + \sum_t  \l^{\textrm{ene}}_{r,t} \left[ -(\eta^{\textrm{dis}}_r)^{-1} g^{\textrm{dis}}_{r,t}+ \eta^{\textrm{sto}}_r  g^{\textrm{sto}}_{r,t} -g^{\textrm{ene}}_{r,t} + \eta^{\textrm{ene}}_r g^{\textrm{ene}}_{r,t-1} \right] \nonumber \\
      = & \sum_t \l_t \left[ g^{\textrm{dis}}_{r,t} - g^{\textrm{sto}}_{r,t}  \right]
\end{align}
The first equality is stationarity for $G^\circ_r$; the second is complimentarity for constraint \eqref{eq:storupper}; the third is stationarity for $g^\circ_{r,t}$ and complimentarity for constraint \eqref{eq:storlower}; the fouth rearranges terms and shifts the cyclic sum over $g^{\textrm{ene}}_{r,t}$; the final equality uses the state of charge constraint \eqref{eq:storsoc}.

The final results shows that the storage recovers its capital costs by arbitrage, charging while prices $\l_t$ are low, and discharging while prices are high.

The relation between market value and LCOE of generators in the system is not affected by the introduction of storage (although the optimal capacities may change).

\subsection{Multi-node long-term equilibrium with network}\label{sec:network}

For multiple nodes the demand and generator variables gain an extra index for the node $n$ to which they are attached, and a term is added to the objective function for the costs $c_\ell$ of each line capacity $F_\ell$ connecting the nodes:
\begin{equation}
   - \sum_{\ell} c_\ell F_\ell
\end{equation}
The flow can move electricity from one node to the other in each hour $f_{\ell,t}$, so that the nodal balance equation is modified
\begin{align}
    \sum_a d_{n,a,t} - \sum_s g_{n,s,t} =  \sum_\ell K_{n\ell} f_{\ell,t} &\hspace{1cm}\perp \hspace{0.32cm} \l_{n,t} \hspace{0.32cm} \forall n,t  \label{sec:netnodalbalance}
\end{align}
where $K_{n\ell}$ is the incidence matrix for the network. This is Kirchhoff's Current Law (KCL).

There are additional constraints on the flows related to the line capacity
\begin{align}
    f_{\ell,t}- F_\ell \leq  0  &\hspace{0.32cm}\perp \hspace{0.32cm} \bar{\mu}_{\ell,t} \hspace{0.32cm} \forall \ell,t \\
    -f_{\ell,t} -F_\ell \leq  0 & \hspace{0.32cm}\perp \hspace{0.32cm} \ubar{\mu}_{\ell,t} \hspace{0.32cm} \forall \ell,t
\end{align}
and to Kirchhoff's Voltage Law (KVL):
\begin{align}
   \sum_\ell C_{\ell,c} x_\ell f_{\ell,t}  = 0  &\hspace{0.32cm}\perp \hspace{0.32cm} \lambda_{c,t} \hspace{0.32cm} \forall c,t
\end{align}
where $c$ label an independent basis of closed cycles in the network defined by the cycle matrix $C_{\ell,c}$, and $x_\ell$ is the series reactance of the line.

From KKT stationarity we get in addition:
\begin{align}
    \frac{\d \cL}{\d f_{\ell,t}} = 0 &\Rightarrow  \sum_{n}\l_{n,t} K_{n,\ell} + \ubar{\m}_{\ell,t} - \bar{\m}_{\ell,t} -\sum_c\l_{c,t}C_{\ell,c}x_\ell = 0 \nonumber\\
            \frac{\d \cL}{\d F_{\ell}} = 0 &\Rightarrow  c_\ell-\sum_t \bar{\m}_{\ell,t} - \sum_t \ubar{\m}_{\ell,t} = 0
\end{align}
and for complementary slackness:
\begin{align}
   \bar{\m}_{\ell,t} (f_{\ell,t} - F_{\ell}) & = 0 \\
  \ubar{\m}_{\ell,t} (f_{\ell,t} + F_{\ell}) & = 0
\end{align}
The no-profit rule becomes:
\begin{align}
c_\ell F_\ell
  & =\sum_{t}  ( \bar{\m}_{\ell,t} + \ubar{\m}_{\ell,t}) F_\ell \\
  & = \sum_{t}  ( \bar{\m}_{\ell,t} - \ubar{\m}_{\ell,t}) f_{\ell,t} \\
   & = \sum_{t,n}\l_{n,t} K_{n,\ell}f_{\ell,t} -\sum_{t,c}\l_{c,t}C_{\ell,c}x_\ell f_{\ell,t}
\end{align}
The first term is the sum over flows $f_{\ell,t}$ multiplied by the price difference between the connect nodes
$\sum_{t,n}\l_{n,t} K_{n,\ell}$, i.e. the congestion revenue. The second term is a distortion that disappears if KVL is not enforced (i.e. in a transport model with only KCL, it would not appear).

Without KVL total costs still equal total revenue, analogous to \eqref{sec:total_costs}:
\begin{align}
  & \sum_{n,s} c_{n,s} G_{n,s} + \sum_{n,s,t} o_{n,s} g_{n,s,t} + \sum_\ell c_\ell F_\ell \\
  & = \sum_{n,s,t}\l_{n,t} g_{n,s,t}  + \sum_{n,\ell,t}\l_{n,t} K_{n,\ell}f_{\ell,t}\\
  & = \sum_{n,a,t} \l_{n,t} d_{n,a,t}
\end{align}
where we have used \eqref{sec:netnodalbalance}.

\subsection{Non-linear generator cost functions}\label{sec:nonlinear}

Suppose we have non-linear, convex functions for the cost of new capacity $C_s(G_s)$ and operation $O_s(g_{s,t})$:
\begin{equation}
    \max_{d_{a,t}, g_{s,t}, G_s}\left[\sum_{a,t} U_{a,t}(d_{a,t}) -  \sum_s C_s(G_s) - \sum_{s,t} O_s(g_{s,t})\right]
\end{equation}
subject to
\begin{align}
   \sum_a d_{a,t} - \sum_s g_{s,t} & =  0 \hspace{0.34cm}\perp \hspace{0.34cm} \l_t \hspace{0.34cm} \forall t \\
    -g_{s,t} & \leq 0 \hspace{0.34cm}\perp \hspace{0.34cm} \ubar{\mu}_{s,t} \hspace{0.34cm} \forall s,t \\
         g_{s,t} - \bar{g}_{s,t} G_s & \leq 0 \hspace{0.34cm}\perp \hspace{0.34cm} \bar{\mu}_{s,t} \hspace{0.34cm} \forall s,t
\end{align}

Now the relationship between costs and revenue becomes
\begin{align}
  G_sC_s'(G_s) + \sum_{t} g_{s,t} O_s' (g_{s,t})
  & = \sum_{t} \l_t g_{s,t}
\end{align}
This becomes a statement about \emph{marginal profit}, i.e. small additions of capacity or generation will not generate any profit or loss.

Generators with convex cost functions with positive derivatives will make a profit since revenue will be higher than costs. For example, if $C_s(G_s) = c_sG_s^n$ for $n>1$ and there are no operating costs, the revenue will be $nc_sG_s^n$, $n$ times higher than the cost. These generators may however be undercut by other generators in the market with different cost functions.

The effects of support and CO$_2$ policies on market value are unchanged.

\subsection{Equivalent problems without constraints}\label{sec:equivalence}

If we have a generic optimisation problem with variables $x_l,y_m$ of the form
\begin{equation}
  \max_{x_l,y_m}\left[ f(x_l) - \sum_m o_m y_m \right]
\end{equation}
subject to equality and inequality constraints:
\begin{align}
  g_i(x_l,y_m) =  0 \hspace{1cm}\perp \hspace{1cm} \l_i \\
      h_j(x_l,y_m) \leq  0 \hspace{1cm}\perp \hspace{1cm} \m_j \\
      \sum_m c_m y_m \leq K \hspace{1cm}\perp \hspace{1cm} \m \label{eq:final}
\end{align}
then we can prove that at the optimal point the solutions for the KKT variables $\l_l,\m_j$ are identical to the following problem without the final constraint \eqref{eq:final}, where we have fixed $\m$ from the above problem as a constant and lifted the constraint into the objective function:
\begin{equation}
  \max_{x_l,y_m}\left[ f(x_l) - \sum_m (o_m + c_m\m) y_m \right]
\end{equation}
subject to equality ($i$) and inequality ($j$) constraints:
\begin{align}
  g_i(x_l,y_m) =  0 \hspace{1cm}\perp \hspace{1cm} \l_i \\
      h_j(x_l,y_m) \leq  0 \hspace{1cm}\perp \hspace{1cm} \m_j
\end{align}
This holds as long as the maximisation problem is a concave function, the inequality constraints are continuously differentiable convex functions and the equality constraints are affine functions (i.e. as long as the KKT conditions are sufficient for optimality).

The lifting of the constraint into the objective function is a standard Lagrangian relaxation. The proof of equivalence of the KKT variables follows by showing that the KKT conditions are identical. From the first problem the only conditions where the extra constraint is relevant is the stationarity for $y_m$
\begin{equation}
    0 =  \frac{\d \cL}{\d y_m} = - o_m - \sum_i \l_i \frac{\d g_i}{\d y_m} - \sum_j \m_j \frac{\d h_j}{\d y_m} - c_m \m
\end{equation}
This is the same as the stationarity for the second problem, where in the second problem the term $c_m \m$ comes from the objective function rather than the constraint. If the final constraint \eqref{eq:final} is not binding, then $\m=0$ by complimentarity and the problems are also identical.  QED.

The values of $x_l,y_m$ are not necessarily identical, but in the case of power system problems, they often are. Since the KKT multipliers for generator constraints $\ubar{\mu}_{s,t},\bar{\mu}_{s,t}$ are identical in both problems, then generators at their upper or lower limits in the first problem are also at their limits in the second problem. The only ambiguities occur for the dispatch of generators that are neither at their lower nor at their upper limits. Where there are multiple generators setting the price with the same linear marginal cost functions, there can be multiple solutions for the same set of KKT multipliers. This is related to the fact that the constant $K$ has disappeared from the second problem. For example, if $K$ is a carbon dioxide budget, then the corresponding carbon tax $\mu$ might not result in a unique generator dispatch if there are many generators with the same marginal costs once the tax is included.

\subsection{Supporting one group of technologies is equivalent to taxing the others}\label{sec:lift}

In this section it is shown that a support policy for technologies $s\in S$ with FiP $\m_\G$ is exactly equivalent to a tax of $\m_\G$ on technologies outside this group $s\notin S$ when demand is perfectly price-inelastic. Switching from forcing in technologies $s\in S$ to forcing out technologies $s\notin S$ results in the prices at each time being lifted by the same constant $\m_\G$. As a result, each market value is also lifted by $\m_\G$.

If $S$ represents VRE technologies, then subsidising VRE is equivalent to taxing non-VRE. For the case that all non-VRE technologies have the same emissions factor, taxing non-VRE is equivalent to a CO$_2$ tax. This latter situation would be relevant for a market with VRE and a single type of gas generator.

For the proof we follow the description of the support policy in Section \ref{sec:retproof}, with the difference that we now assume that the demand is perfectly price-inelastic in the model, i.e. that the variables $d_{a,t}$ are constants.

If a subset of generators $S$ is singled out and forced to meet at least a fraction $\gamma\in [0,1]$ of the total demand $\sum_{a,t,} d_{a,t}$, this is represented with the constraint from equation \eqref{eq:vre}
\begin{equation}
  \sum_{s\in S,t} g_{s,t} \geq \G = \g\sum_{a,t} d_{a,t} \hspace{0.34cm}\perp \hspace{0.34cm} \m_{\G} \label{ref:proVRE}
\end{equation}
For generators included in the constraint, $s\in S$, stationarity is
\begin{align}
    \frac{\d \cL}{\d g_{s,t}} = 0 & \Rightarrow  \l_t = o_{s} - \ubar{\m}_{s,t} + \bar{\m}_{s,t} - \m_{\Gamma} \label{eq:sbefore}
\end{align}
and for other generators $s\notin S$ we have as in  equation \eqref{eq:statvanilla}
\begin{align}
    \frac{\d \cL}{\d g_{s,t}} = 0 & \Rightarrow  \l_t = o_{s} - \ubar{\m}_{s,t} + \bar{\m}_{s,t}  \label{eq:notsbefore}
\end{align}

Now suppose we manipulate the constraint \eqref{ref:proVRE} by subtracting both sides from the total generation:
\begin{equation}
 \sum_{s,t} g_{s,t} -  \sum_{s\in S,t} g_{s,t} \leq \sum_{s,t} g_{s,t} - \g\sum_{a,t} d_{a,t} \hspace{0.34cm}\perp \hspace{0.34cm} \m_{\G}
\end{equation}
Since we subtracted \eqref{ref:proVRE}, the direction of the inequality reverses. Now rearrange, remembering that we are in a lossless system with balanced demand $\sum_{s,t} g_{s,t} = \sum_{a,t} d_{a,t}$ from \eqref{eq:balance}:
\begin{equation}
 \sum_{s\notin S,t} g_{s,t}  \leq (1 - \g)\sum_{a,t} d_{a,t} \hspace{0.34cm}\perp \hspace{0.34cm} \m_{\G}
\end{equation}
This will give exactly the same results as the previous problem, since the constraint is identical up to a sign and a constant factor added to both sides, except now we are restricting technologies not in $S$ to be no more than a fraction $(1-\g)$ of the demand, and the prices $\l_t$ are lifted by $\m_\G$.

To show the effect on prices, consider stationarity for $s\in S$:
\begin{align}
    \frac{\d \cL}{\d g_{s,t}} = 0 & \Rightarrow  \l_t = o_{s} - \ubar{\m}_{s,t} + \bar{\m}_{s,t}  \label{eq:safter}
\end{align}
and for $s\notin S$:
\begin{align}
    \frac{\d \cL}{\d g_{s,t}} = 0 & \Rightarrow  \l_t  = o_s - \ubar{\m}_{s,t} + \bar{\m}_{s,t} + \m_{\Gamma}  \label{eq:notsafter}
\end{align}

By comparing equations \ref{eq:sbefore} and  \ref{eq:notsbefore} for the support policy for $S$ to the equivalent equations \ref{eq:safter} and \ref{eq:notsafter}, you can see that a subsidy for $S$ with FiP $\m_\G$ is equivalent to a tax on non-$S$ of $\m_\G$. Switching from the subsidy to the tax results in the lift of all prices by a constant factor $\l_t \mapsto \l_t + \m_\G$.

From an `effective bid' perspective, forcing in the share of $S$ subtracts $\m_\G$ from $S$ marginal cost bids, while forcing out the non-$S$ share adds $\m_\G$ to the non-$S$ bids, simply lifting all bids on the merit-order curve by $\m_\G$.

Market value is also just lifted by $\m_\Gamma$ for all technologies $s$
\begin{equation}
  MV_s \equiv \frac{\sum_t g_{s,t} \l_t}{\sum_t g_{s,t}} \mapsto \frac{\sum_t g_{s,t}( \l_t + \m_\G)}{\sum_t g_{s,t}} = MV_s + \mu_\G
\end{equation}

\section{Technology assumptions}\label{sec:assumptions}

The technology assumptions from the original model EMMA \cite{HIRTH2013218} and our model PyPSA are compared in Table \ref{tab:assumptions}. While nuclear and lignite with CCS are disabled in the main calculations, for the calculations in the Appendix with nuclear the costs from \cite{schroeder2013} are applied, to reflect experience in recent projects.

Power plant lifetimes are taken from \cite{HIRTH2013218} (nuclear has a lifetime of 50 years, while other plants have 25 years).

Battery assumptions are drawn from \cite{budischak2013}, hydrogen (H$_2$) electrolysis from \cite{SCHMIDT201730470} and underground H2 storage from \cite{nrel2009}.

The costs of transmission expansion between the countries are derived following \cite{Schlachtberger2017}, assuming high voltage alternating current connections, that transmission covers the distance between the geographical mid-points of the countries with 25\% extra length to account for non-direct routes, a 33\% capacity buffer for $N-1$ failures and reactive power flows, and a 40 year lifetime for the new transmission assets.

\begin{table}[t]
  \centering
  \setlength{\tabcolsep}{6pt}
  \begin{tabular}{@{} llrr @{}}
    \toprule
    {\bf Quantity} & {\bf Unit} & {\bf EMMA}  & {\bf PyPSA}  \\
    \midrule
    wind cost & \euro/kW & 1300 & 1040 \\
    solar cost & \euro/kW & 2000 & 510 \\
    nuclear cost & \euro/kW & 4000 & 6000 \\
    nuclear fuel cost & \euro/MWh\th & 3 & 3 \\
    lignite cost & \euro/kW & 2200 & 2200 \\
    lignite fuel cost & \euro/MWh\th & 3 & 3 \\
    lignite+CCS cost & \euro/kW & 3500 &  n/a \\
    lignite+CCS fuel cost & \euro/MWh\th & 3 & n/a \\
    coal cost & \euro/kW & 1500 & 1500 \\
    coal fuel cost & \euro/MWh\th & 11.5 & 11.5 \\
    CCGT cost & \euro/kW & 1000 & 1000 \\
    CCGT fuel cost & \euro/MWh\th & 25 & 25 \\
    OCGT cost & \euro/kW & 600 & 600 \\
    OCGT fuel cost & \euro/MWh\th & 50 & 50 \\
    load shedding cost & \euro/MWh\el & 1000 & 1000 \\
    battery inverter & \euro/kW & n/a & 333 \\
    battery storage & \euro/kWh & n/a & 167 \\
    H$_2$ electrolysis & \euro/kW\el & n/a &  750 \\
    H$_2$ electrolysis efficiency & \% & n/a & 80 \\
    H$_2$ turbine & \euro/kW\el & n/a & 800 \\
    H$_2$ storage & \euro/kWh & n/a & 0.5 \\
    transmission expansion & \euro/(MWkm) & n/a & 400 \\
    \bottomrule
  \end{tabular}
  \caption{Comparison of technology assumptions in the different models.}
  \label{tab:assumptions}
\end{table}

\section{Comparison of PyPSA to EMMA results for RMV}\label{sec:validation}

\begin{figure}[!t]
\centering
    \includegraphics[trim=0 0cm 0 0cm,width=\linewidth,clip=true]{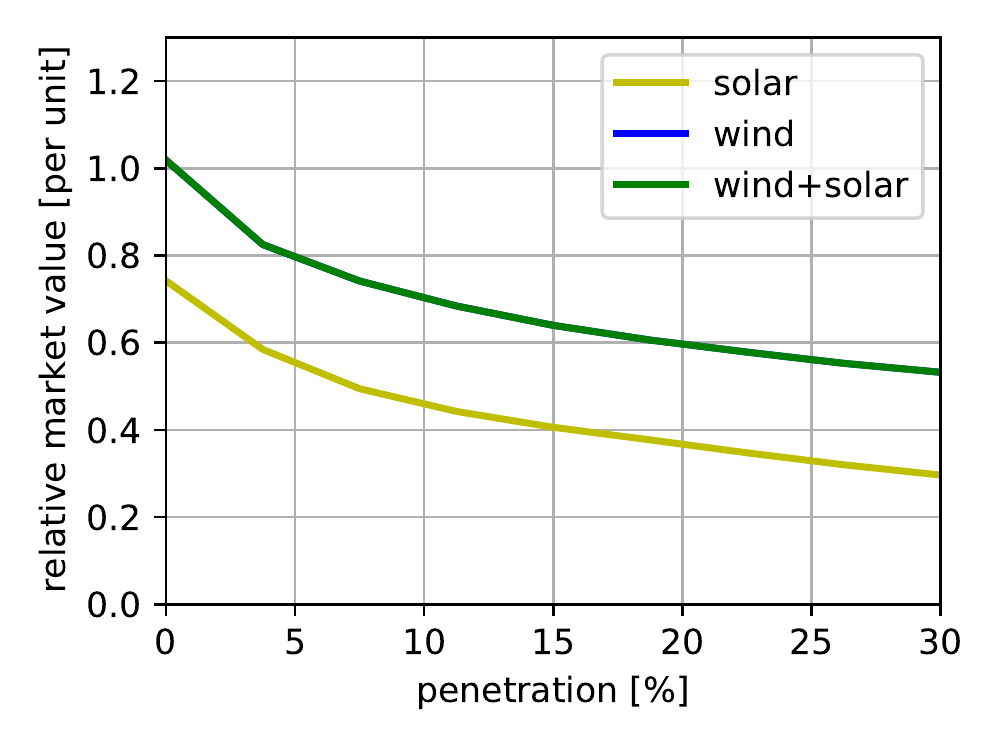}
\caption{The relative market value of wind and solar as the share of their available energy is changed in the model for the case without storage or transmission reinforcement using all costs from \cite{HIRTH2013218}.}
\label{fig:rmv-no_storage-vanilla}
\end{figure}

\begin{figure}[!t]
\centering
    \includegraphics[trim=0 0cm 0 0cm,width=\linewidth,clip=true]{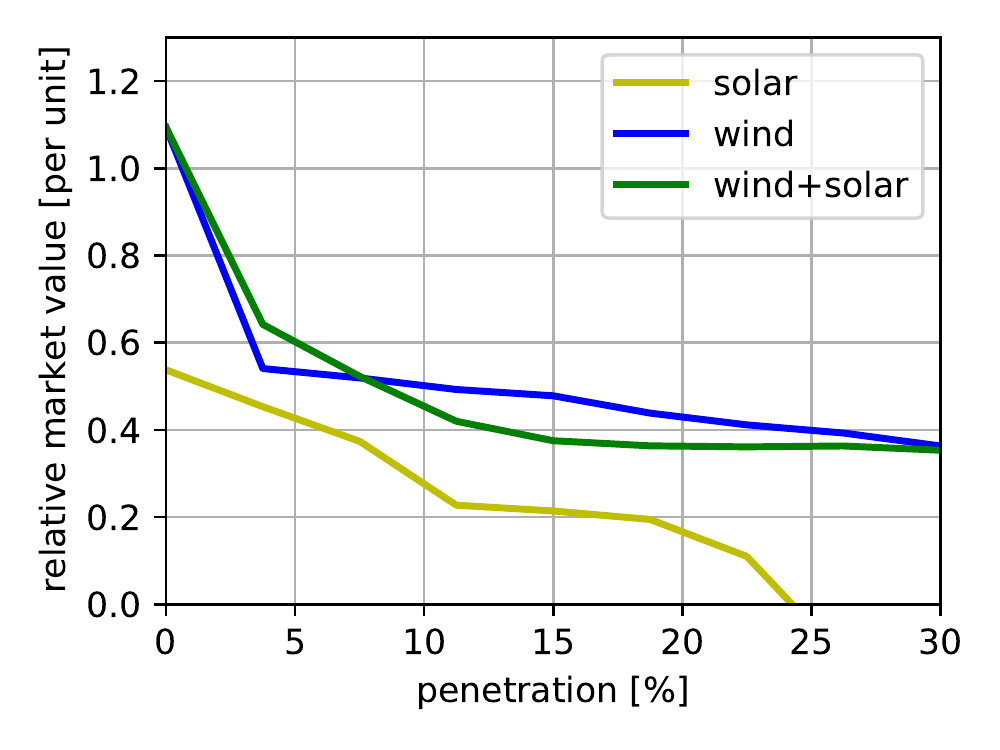}
\caption{The relative market value of wind and solar as the share of their dispatched energy is changed in the model for the case without storage or transmission reinforcement using our updated costs.}
\label{fig:rmv-no_storage-new}
\end{figure}

In this section the results from \cite{HIRTH2013218} for the long-term relative market values of solar and wind are compared to the results from the
reimplementation in PyPSA. Figure \ref{fig:rmv-no_storage-vanilla} shows the relative market values in the PyPSA model with the same technology assumptions
as \cite{HIRTH2013218} and with a constraint on available renewable energy following \cite{HIRTH2013218} (see \ref{sec:retcapproof}).
Figure \ref{fig:rmv-no_storage-new} shows the results with the wind and solar costs updated, the CO$_2$ price set to zero, the removal of nuclear and CCS
as options, and a constraint on dispatched renewable energy following Section \ref{sec:retproof}. Table \ref{tab:rmvs} compares the relative market values
for different sampling points.

First we compare the results from EMMA (column 1 in Table \ref{tab:rmvs} and Figures 18 and 27 in  \cite{HIRTH2013218}) and the reimplementation in PyPSA
(column 2 in Table \ref{tab:rmvs} and Figure \ref{fig:rmv-no_storage-vanilla}). While there is clear agreement in the overall shape and trajectory of the
curves, in three of the four cases PyPSA underestimates the relative market values compared to EMMA, particularly for the case of solar at 15\%
penetration. There are several factors causing the disagreement between EMMA and PyPSA: EMMA has baseload incentives which alter prices; EMMA has
incentives for flexible generators like OCGT that reduce their capital costs, encouraging higher marginal cost generators into the market and pushing up
prices; for the denominator of the RMV, EMMA takes a simple price average over time, while PyPSA load-weights average prices over time (emphasising times
of high load when prices are either higher (evening) or lower (midday solar peak)); and finally it appears the EMMA code uses a lifetime of 25 years for
nuclear rather than the 50 years applied here. The solar disagreement is also large because the slope of the curve here is steep, so any deviation is
magnified.

%XX solar reaches 0.58 at x\% rather than 15\%

Next we compare the results from  the reimplementation in PyPSA (column 2 in Table \ref{tab:rmvs} and Figure \ref{fig:rmv-no_storage-vanilla}) and the
version of PyPSA with updated assumptions (column 3 in Table \ref{tab:rmvs} and Figure \ref{fig:rmv-no_storage-new}). One of
the main change in costs was a reduction in solar costs, and this is reflected in Figures \ref{fig:rmv-no_storage-vanilla} and \ref{fig:rmv-no_storage-new}
by the fact that the case with both wind and solar now differs from the pure wind case, since solar is competitive. The reduction of the CO$_2$ price from
20~\euro/tCO$_2$ to zero helps to suppress prices. And finally the constraint on dispatched energy rather than available energy means that costs decrease
faster and go to zero, since VRE subsidy can cause hours of negative prices. When we constrain available energy, VRE is curtailed at zero price, meaning
that RMV flatlines at high penetrations as in Figure \ref{fig:rmv-no_storage-vanilla} (note that at high penetrations, a lot of the energy is
curtailed).

\begin{table}[t]
  \centering
  \setlength{\tabcolsep}{6pt}
  \begin{tabular}{@{} lrrr @{}}
    \toprule
    {\bf Model} & {\bf EMMA} & {\bf PyPSA}  & {\bf PyPSA}  \\
    {\bf Costs} & \cite{HIRTH2013218} & \cite{HIRTH2013218} & {\bf adjusted} \\
    \midrule
    solar at 0\% & 0.9 & 0.74 & 0.54 \\
    solar at 15\% & 0.58 & 0.41 & 0.21 \\
    wind at 0\% & 1.1 & 1.02 & 1.1 \\
    wind at 30\% & 0.64 & 0.53 & 0.36 \\
    \bottomrule
  \end{tabular}
  \caption{Comparison of relative market values in the different models.}
  \label{tab:rmvs}
\end{table}

\section{Additional results}\label{app:results}

\subsection{Support policies for wind, solar and nuclear separately}\label{app:ret}

\begin{figure}[!t]
\centering
    \includegraphics[trim=0 0cm 0 0cm,width=\linewidth,clip=true]{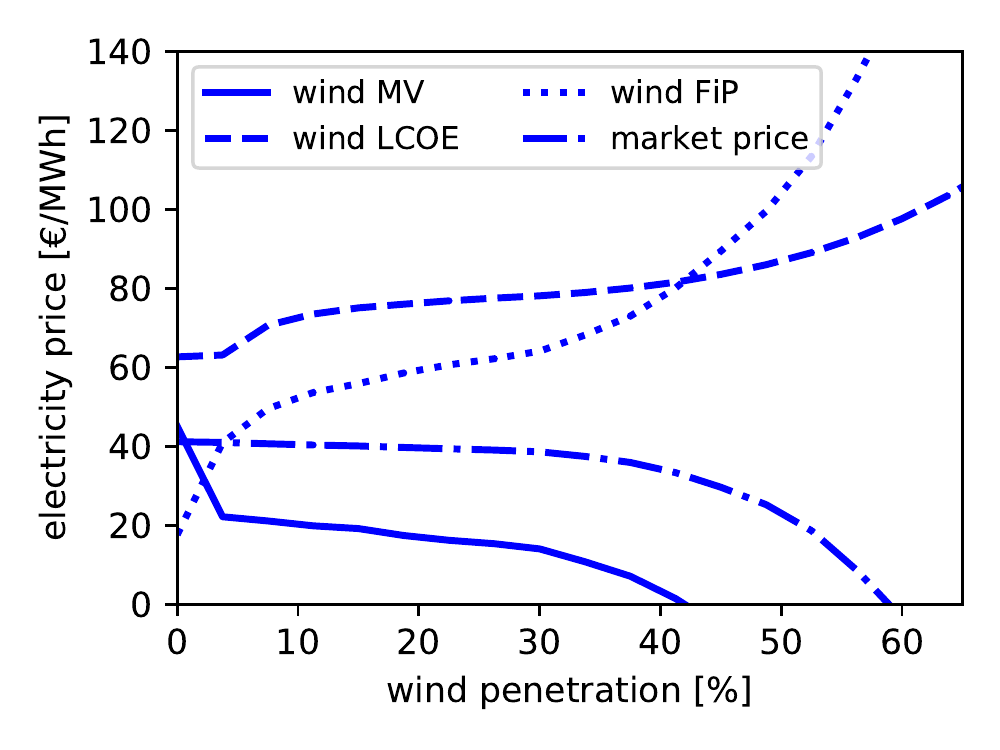}
\caption{Market quantities as the penetration of wind energy covering electricity demand is increased. In this case there is no additional flexibility from storage or transmission reinforcement.}
\label{fig:mwh-wind}
\end{figure}

\begin{figure}[!t]
\centering
    \includegraphics[trim=0 0cm 0 0cm,width=\linewidth,clip=true]{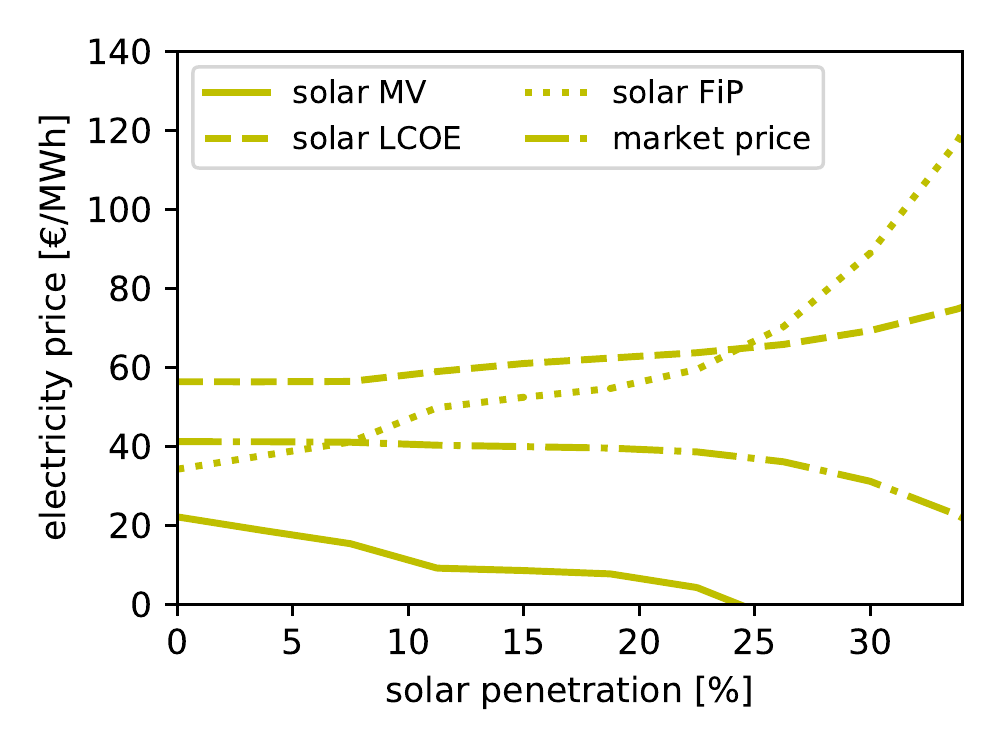}
\caption{Market quantities as the penetration of solar energy covering electricity demand is increased. In this case there is no additional flexibility from storage or transmission reinforcement.}
\label{fig:mwh-solar}
\end{figure}

\begin{figure}[!t]
\centering
    \includegraphics[trim=0 0cm 0 0cm,width=\linewidth,clip=true]{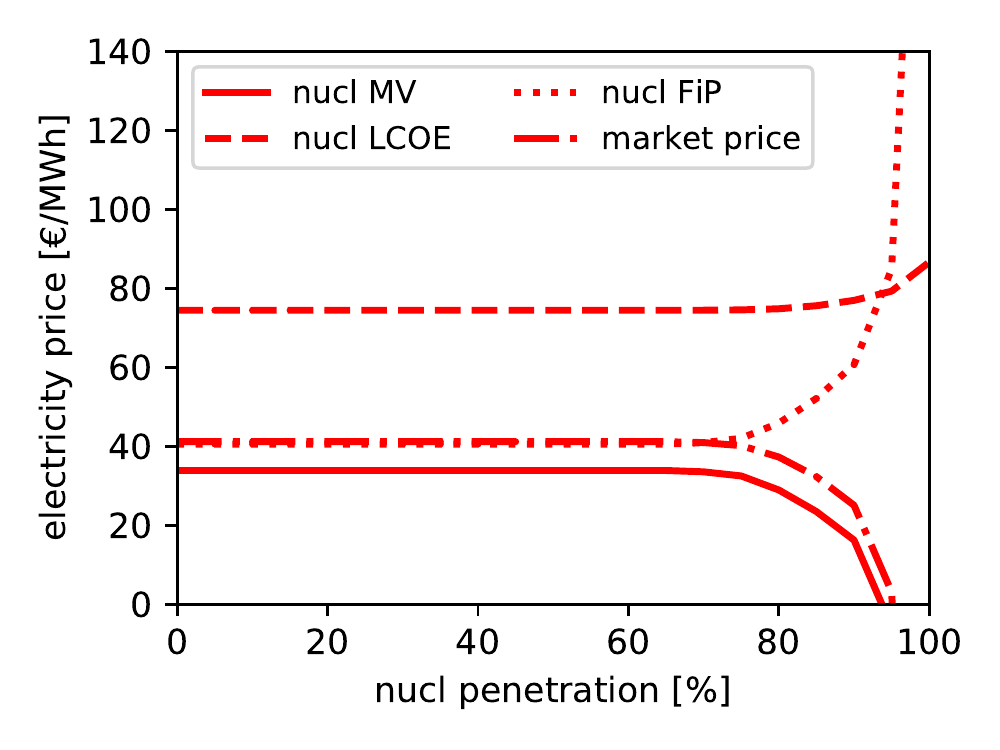}
\caption{Market quantities as the penetration of nuclear energy covering electricity demand is increased. In this case there is no additional flexibility from storage or transmission reinforcement.}
\label{fig:mwh-nuclear}
\end{figure}

Results for support policies applied separately to wind are shown in Figure  \ref{fig:mwh-wind}, solar in Figure \ref{fig:mwh-solar} and nuclear in \ref{fig:mwh-nuclear}.

The solar market value declines much faster than for wind, as has been seen in previous results in this paper and elsewhere in the literature.

The effect of forced penetration on nuclear is similar. The equilibrium solution without the constraint does not contain nuclear, because the cost is too high; a non-zero subsidy is required to cover the difference between its average market value and the LCOE.
Because it is available at all times, it achieves penetration of up to 75\% before the market value declines, which corresponds to the minimum value of the load. Above this point, it reaches lower capacity factors, forcing the LCOE up and the market value down.

\subsection{Support policies with no negative prices}\label{app:ret-zero}

\begin{figure}[!t]
\centering
    \includegraphics[trim=0 0cm 0 0cm,width=\linewidth,clip=true]{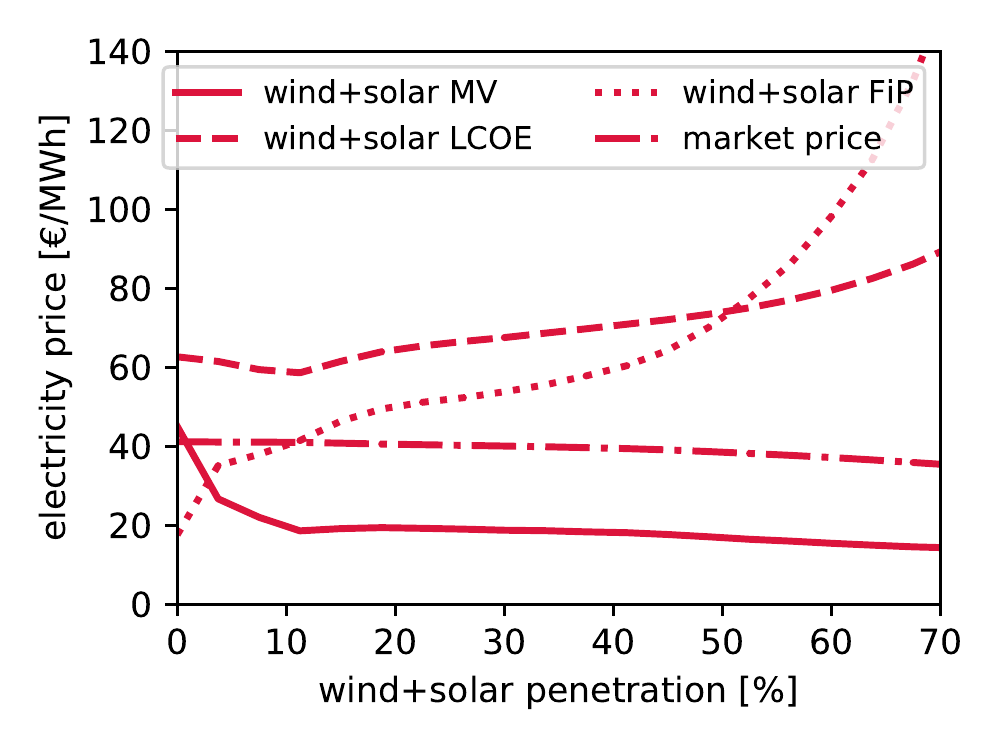}
\caption{Market quantities for a VRE support policy for wind and solar with no negative prices and without additional flexibility.}
\label{fig:mwh-zeroed-wind-solar}
\end{figure}

\begin{figure}[!t]
\centering
    \includegraphics[trim=0 0cm 0 0cm,width=\linewidth,clip=true]{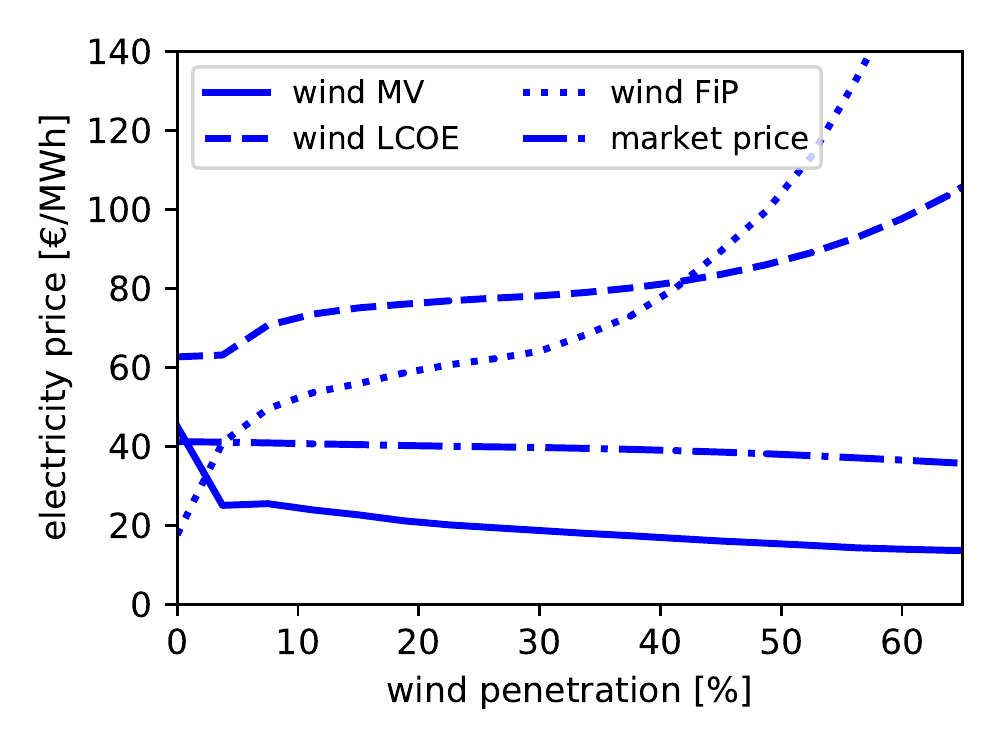}
\caption{Market quantities for a VRE support policy for wind with no negative prices and without additional flexibility.}
\label{fig:mwh-zeroed-wind}
\end{figure}

\begin{figure}[!t]
\centering
    \includegraphics[trim=0 0cm 0 0cm,width=\linewidth,clip=true]{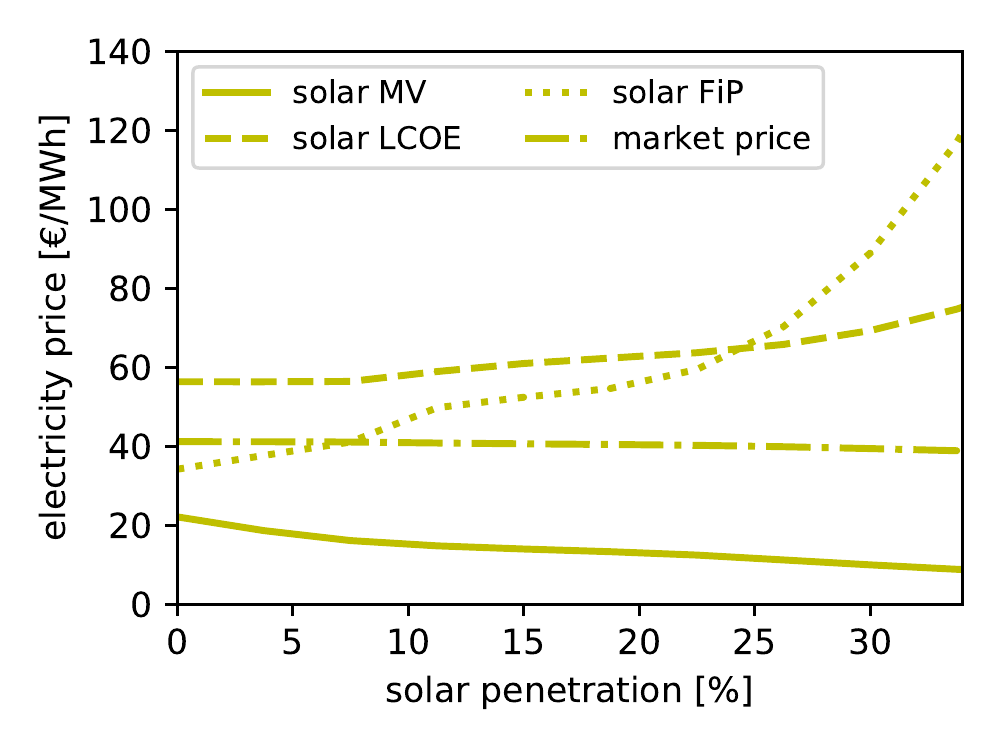}
\caption{Market quantities for a VRE support policy for solar with no negative prices and without additional flexibility.}
\label{fig:mwh-zeroed-solar}
\end{figure}

\begin{figure}[!t]
\centering
    \includegraphics[trim=0 0cm 0 0cm,width=\linewidth,clip=true]{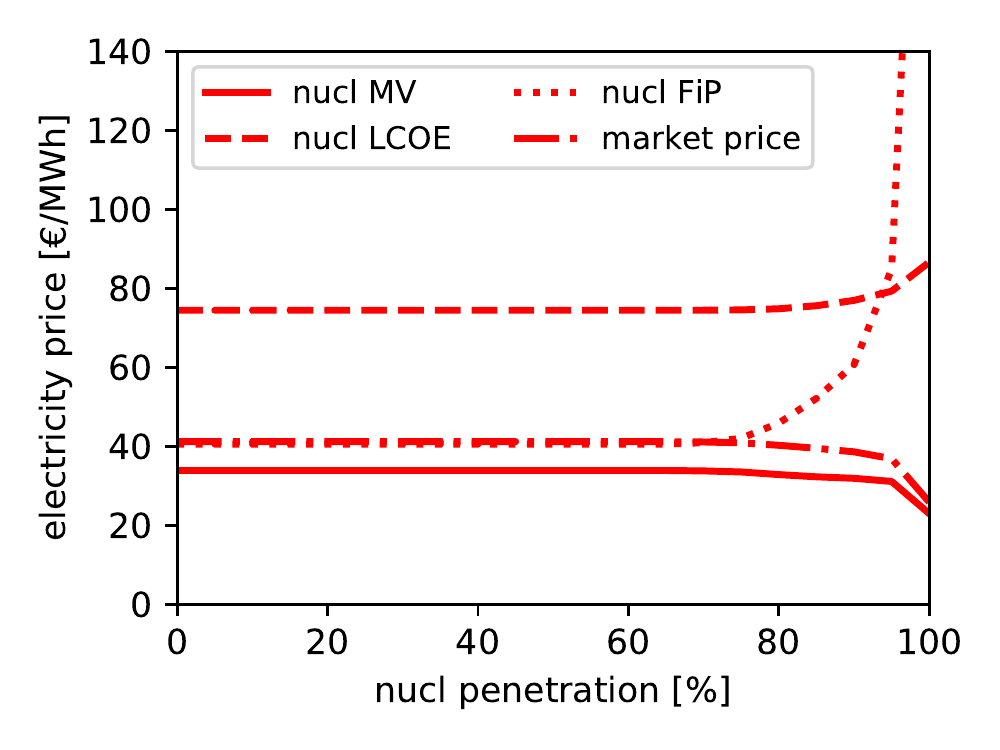}
\caption{Market quantities for a VRE support policy for nuclear with no negative prices and without additional flexibility.}
\label{fig:mwh-zeroed-nuclear}
\end{figure}

In this section we take the results from the support policy and forbid negative prices, by setting the price to zero whenever it goes below zero. The results for systems without additional flexibility are shown for wind and solar in Figure \ref{fig:mwh-zeroed-wind-solar}, for wind in Figure \ref{fig:mwh-zeroed-wind}, for solar in Figure \ref{fig:mwh-zeroed-solar} and for nuclear in Figure \ref{fig:mwh-zeroed-nuclear}. In all cases the average market price falls only gradually. The market values still decline well below the point of cost recovery, but the decline is gentler than when negative prices are allowed, and the market values never turn negative.

\subsection{CO$_2$ policy details}\label{app:co2}

\begin{figure}[!t]
\centering
    \includegraphics[trim=0 0cm 0 0cm,width=\linewidth,clip=true]{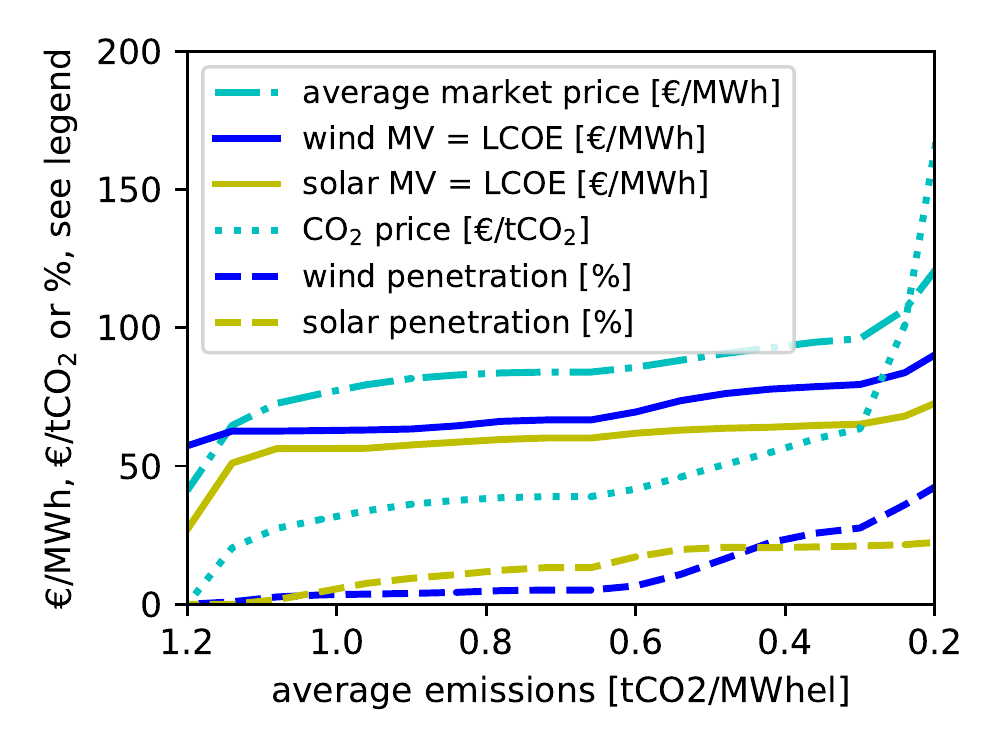}
\caption{Market quantities as the average CO$_2$ emission factor is reduced to zero for a scenario without additional flexibility.}
\label{fig:mwh-co2}
\end{figure}

In Figure \ref{fig:mwh-co2} the effect of a CO$_2$ constraint on average market prices, wind and solar MV (equal to LCOE), the CO$_2$ dual price and the wind and solar penetrations are plotted. From the unconstrained equilibrium with emissions of around 1.2~tCO$_2$/MWh\el down to about 0.7~tCO$_2$/MWh\el, emissions are reduced by substituting coal for lignite, and gas for goal. Below 0.7~tCO$_2$/MWh\el, wind and solar penetrations rise steadily to replace natural gas. The CO$_2$ price required to reach each target rises in steps as particular fuels are substituted, before rising very steeply below 0.3~tCO$_2$/MWh\el, where it gets harder to match the variable profiles of wind and solar with the load.

The market price increases with a stricter CO$_2$ constraint, since the rising CO$_2$ price increases all effective marginal costs $o_s \to o_s + e_s \mu_K$. The market values of wind and solar initially remain steady, since they are equal to the LCOE, which is stable. However, as penetration rises, curtailment increases and the LCOE drops.

\begin{figure}[!t]
\centering
    \includegraphics[trim=0 0cm 0 0cm,width=\linewidth,clip=true]{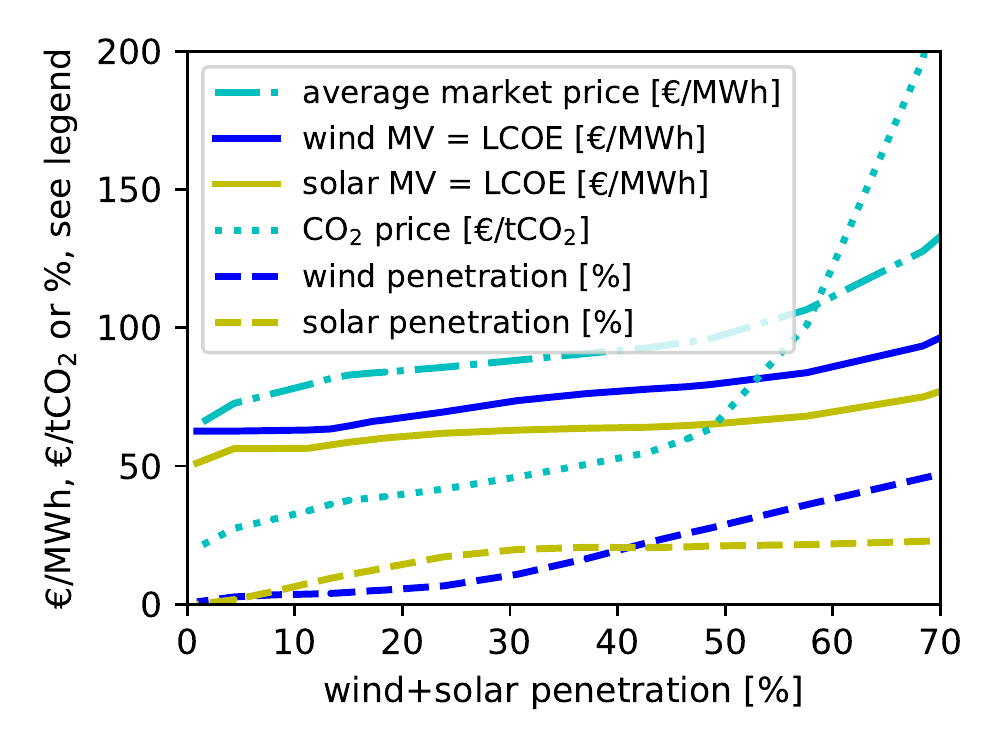}
\caption{Market quantities as the average CO$_2$ emission factor is reduced to zero for a scenario without additional flexibility.}
\label{fig:mwh-pen-co2}
\end{figure}

Figure \ref{fig:mwh-pen-co2} shows the corresponding figure as a function of the combined wind and solar penetration.

\begin{figure}[!t]
\centering
    \includegraphics[trim=0 0cm 0 0cm,width=\linewidth,clip=true]{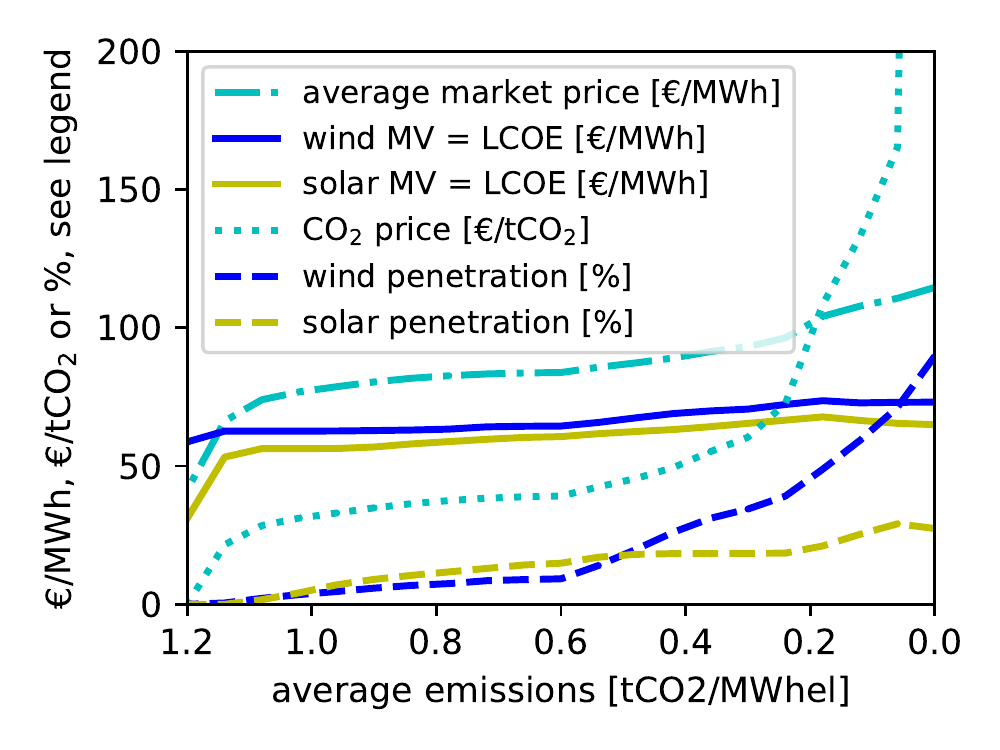}
\caption{Market quantities as the average CO$_2$ emission factor is reduced to zero for a scenario with transmission expansion as well as short- and long-term storage.}
\label{fig:mwh-co2-storage-trans}
\end{figure}

\begin{figure}[!t]
\centering
    \includegraphics[trim=0 0cm 0 0cm,width=\linewidth,clip=true]{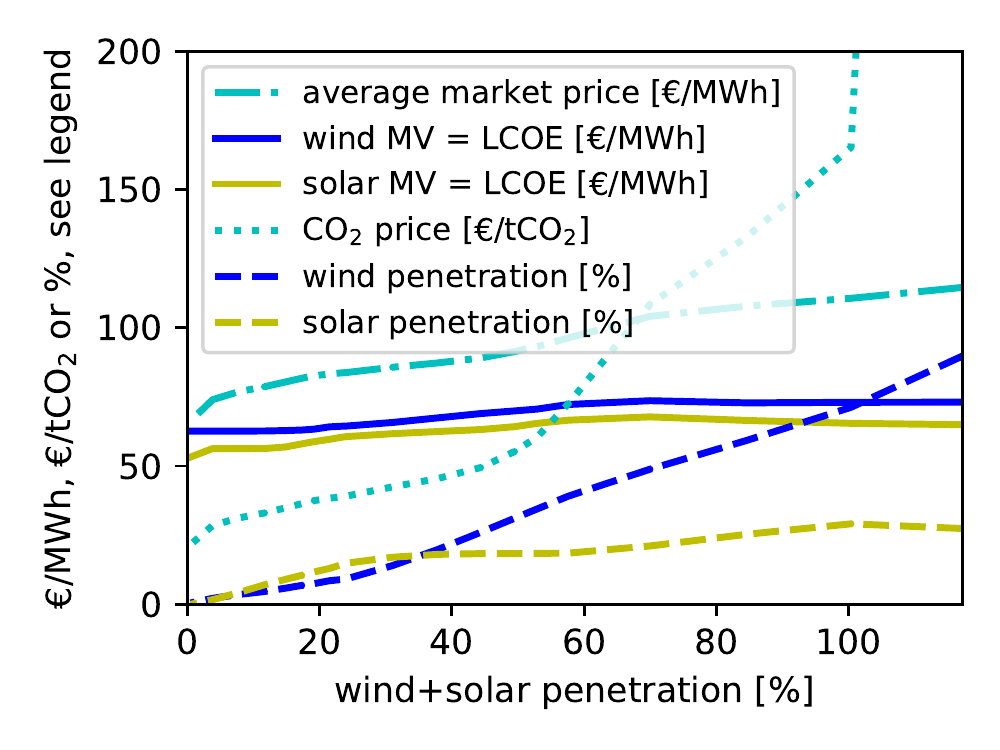}
\caption{Market quantities as the average CO$_2$ emission factor is reduced to zero for a scenario with transmission expansion as well as short- and long-term storage.}
\label{fig:mwh-pen-co2-storage-trans}
\end{figure}

The analysis was repeated with the addition of transmission expansion and storage investment possibilities for battery and hydrogen storage; the corresponding results are plotted in Figure \ref{fig:mwh-co2-storage-trans} as a function of the CO$_2$ limit and in Figure \ref{fig:mwh-pen-co2-storage-trans} as a function of penetration. With the additional flexibility, curtailment is limited and a VRE penetration of 100\% without either the average market price (which reflects total system cost) or the market values rising drastically.

\begin{figure}[!t]
\centering
    \includegraphics[trim=0 0cm 0 0cm,width=\linewidth,clip=true]{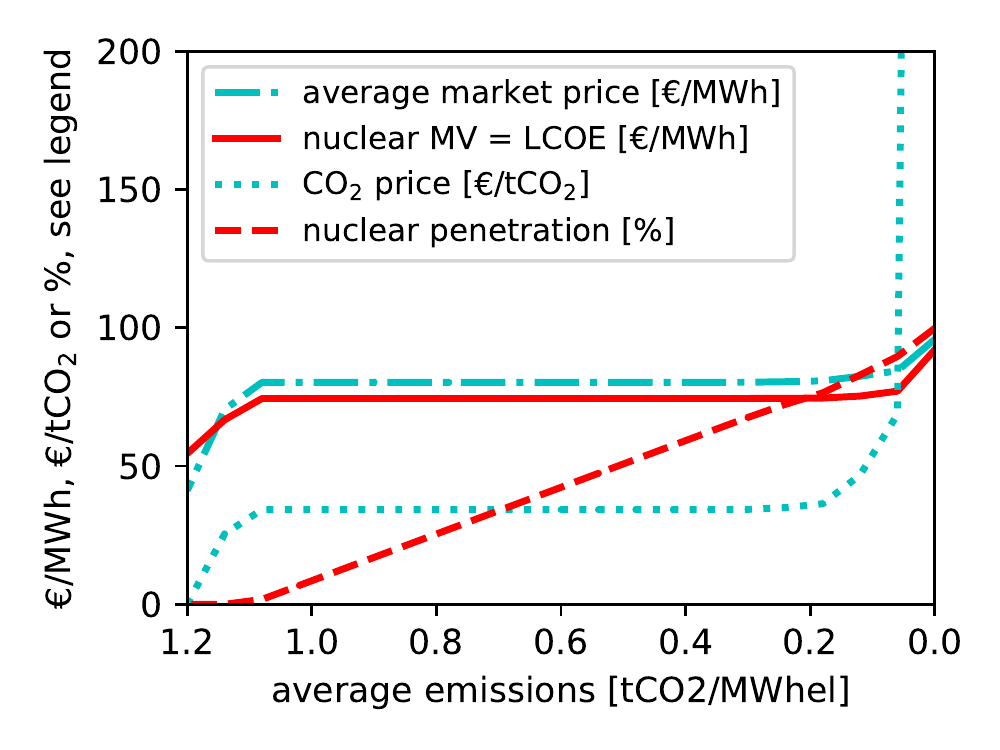}
\caption{Market quantities as the average CO$_2$ emission factor is reduced to zero for a scenario with nuclear and no wind or solar.}
\label{fig:mwh-co2-nuclear}
\end{figure}

\begin{figure}[!t]
\centering
    \includegraphics[trim=0 0cm 0 0cm,width=\linewidth,clip=true]{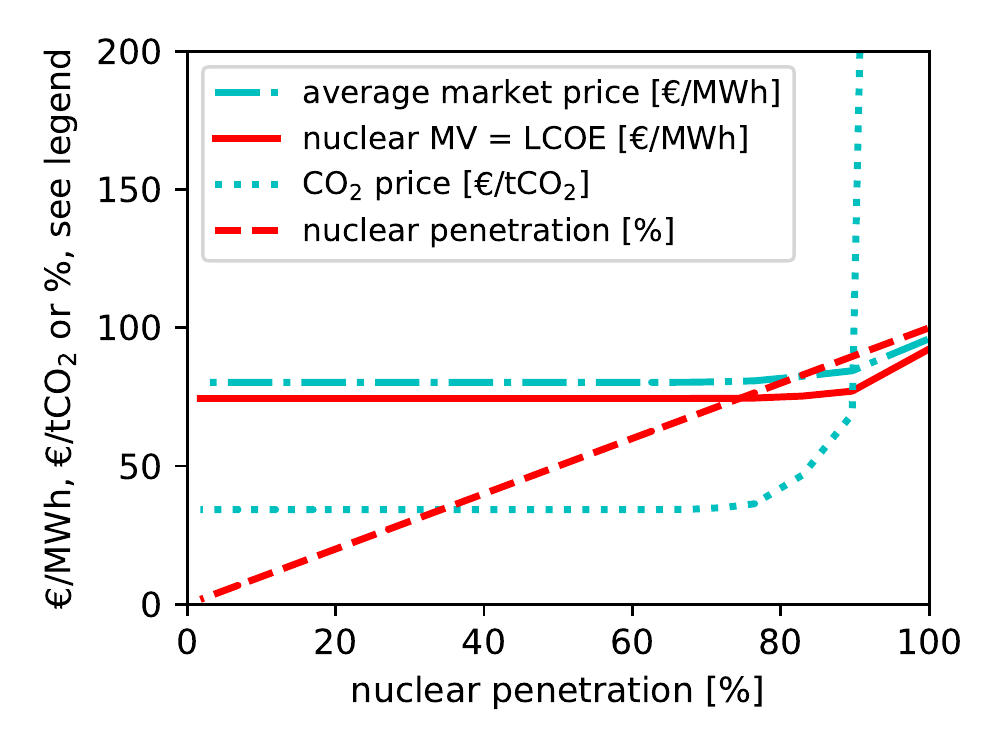}
\caption{Market quantities as the average CO$_2$ emission factor is reduced to zero for a scenario with nuclear and no wind or solar.}
\label{fig:mwh-pen-nuclear}
\end{figure}

In Figures \ref{fig:mwh-co2-nuclear} and \ref{fig:mwh-pen-nuclear} the analysis is reproduced for a scenario where the CO$_2$ limit is brought down with only nuclear as a low CO$_2$ technology. Once the CO$_2$ price reaches 34~\euro/tCO$_2$, nuclear is competitive with lignite and rises to a share of 68\% of electricity generation. To reach 90\% penetration, the CO$_2$ price must rise to 69~\euro/tCO$_2$.

\subsection{Relative market value}

\begin{figure}[!t]
\centering
    \includegraphics[trim=0 0cm 0 0cm,width=\linewidth,clip=true]{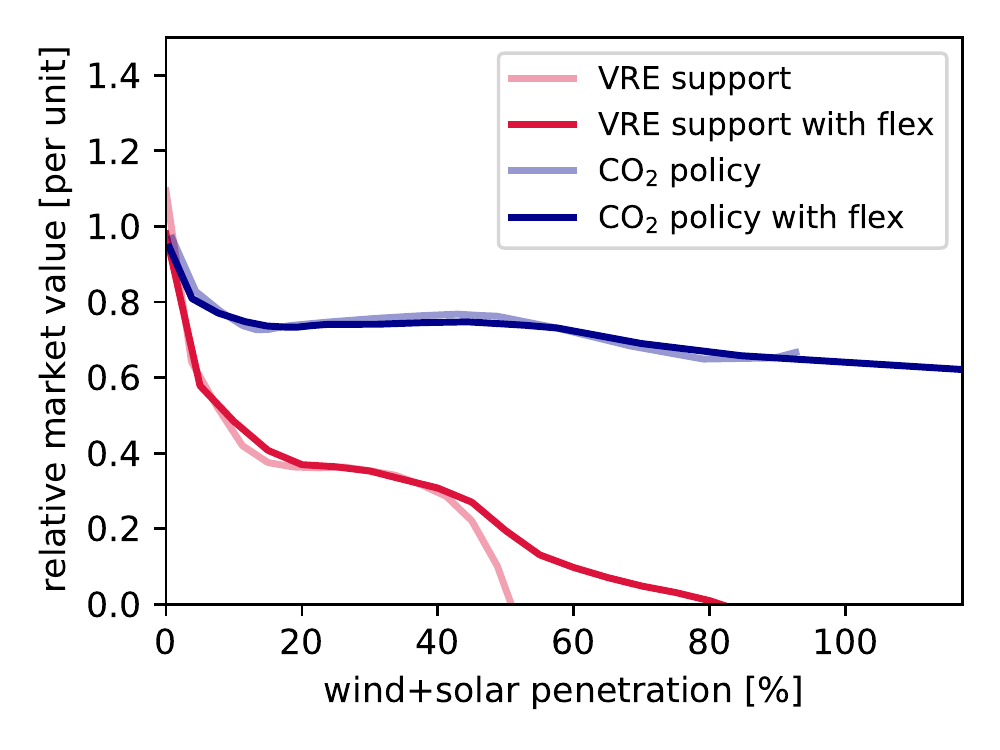}
\caption{Relative market values (RMV) for VRE support versus CO$_2$ policies with and without flexibility.}
\label{fig:comparison-rmv}
\end{figure}

The relative market value (RMV), also called the value factor in \cite{HIRTH2013218}, is the ratio of the market value to the load-weighted average market price, see equations \eqref{eq:rmvdef} and \eqref{eq:rmvid}. The absolute market values from Figure \ref{fig:co2-compare} are shown as relative market values in Figure \ref{fig:comparison-rmv}. Under a VRE support policy, the RMV still goes to zero and then negative. The RMV under a CO$_2$ policy shows a shallow decline, which can be explained from equation \eqref{eq:rmvid}. Equation \eqref{eq:rmvid} shows that the RMV can also be expressed as the ratio between the technology's fraction of system costs to its share of demand. Since at full penetration VRE covers all of the demand and storage losses, the RMV simply reflects the fraction of VRE in the total system costs. The RMV ends up at 0.62, reflecting the fraction of VRE in the system costs from Figure \ref{fig:sys_cost} (remaining costs coming from transmission and storage for balancing).

\subsection{Comparing system cost as function CO$_2$ emissions}\label{app:sys-co2}

\begin{figure}[!t]
\centering
    \includegraphics[trim=0 0cm 0 0cm,width=\linewidth,clip=true]{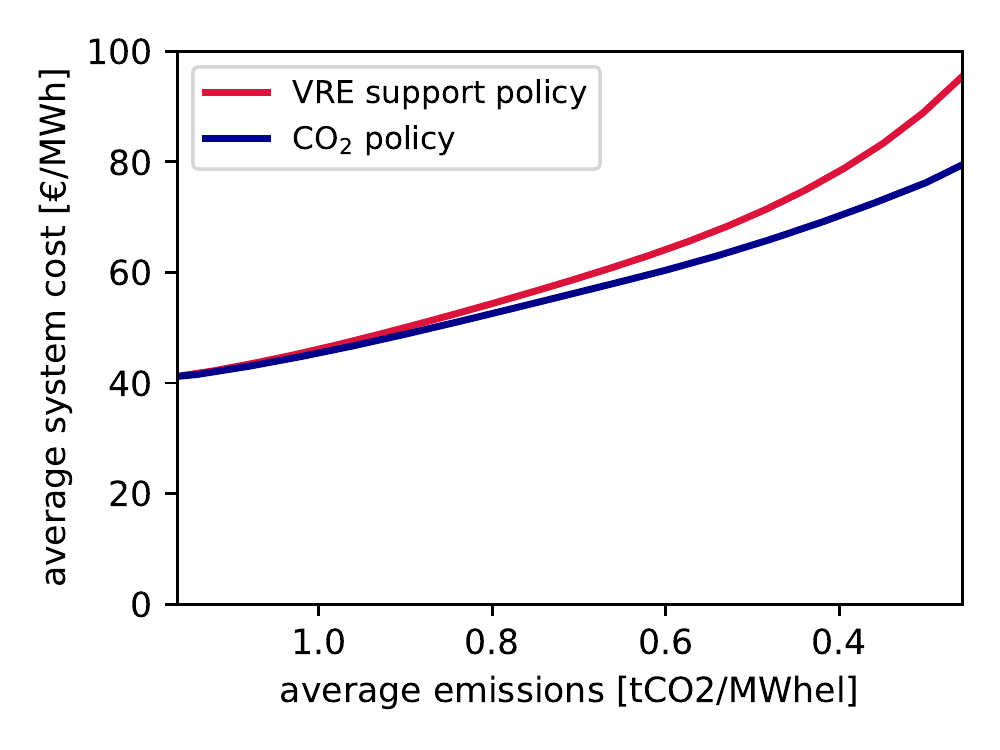}
\caption{Comparison of system costs for VRE support and CO$_2$ policies as a function of average system CO$_2$ emissions, without flexibility.}
\label{fig:sys_cost-comparison-co2}
\end{figure}

In Figure \ref{fig:sys_cost-comparison} we compared the system costs under the VRE and CO$_2$ policies as a function of the penetration of wind and solar without flexibility. In Figure \ref{fig:sys_cost-comparison-co2} we provide the complementary figure comparing the system costs of the two policies as a function of average system CO$_2$ intensity. For the system setup here, the policies provide similar results until higher penetrations, at which the CO$_2$ policy is, unsurprisingly, more efficient at reducing CO$_2$ emissions.

\subsection{Price duration curves for a CO$_2$ policy with flexibility}\label{app:duration}

In this section we discuss the price duration curves in the model as the CO$_2$ budget is reduced to zero for the case of wind and solar in the presence of additional flexibility from storage and transmission reinforcement.

\begin{figure}[!t]
\centering
    \includegraphics[trim=0 0cm 0 0cm,width=\linewidth,clip=true]{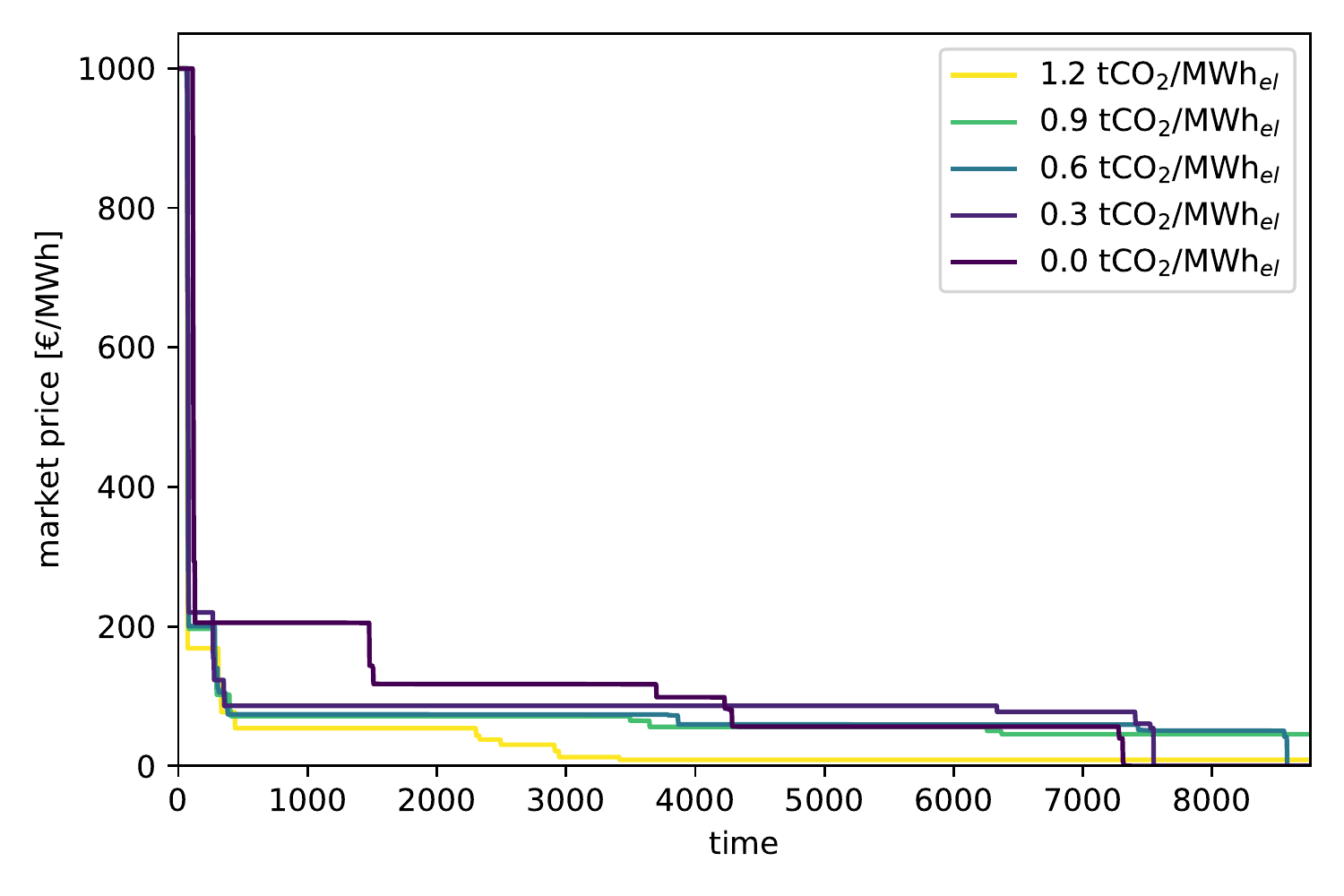}
\caption{Price duration curves for CO$_2$ policies for different average system CO$_2$ emissions, with flexibility from storage and transmission reinforcement.}
\label{fig:price_duration}
\end{figure}

Figure \ref{fig:price_duration} shows the price duration curves for different levels of average system CO$_2$ emissions. For higher emissions, the curve is flatter and there are no times of zero prices. As emissions reduce, there are more hours with zero prices set by wind and solar, rising to 16\% of all hours, and more times with higher prices. When fossil generators are pushed out of the system, arbitrage between storage and transmission sets the non-zero prices, either by demand bids when VRE is abundant or supply bids when VRE is scarce. Note that the system does not degenerate into a singular system where prices are either zero or the value of lost load (1000~\euro/MWh in this case), as is sometimes assumed. Similar price duration curves were observed in a sector-coupled model in \cite{boettger2021}.

\begin{figure}[!t]
\centering
    \includegraphics[trim=0 0cm 0 0cm,width=\linewidth,clip=true]{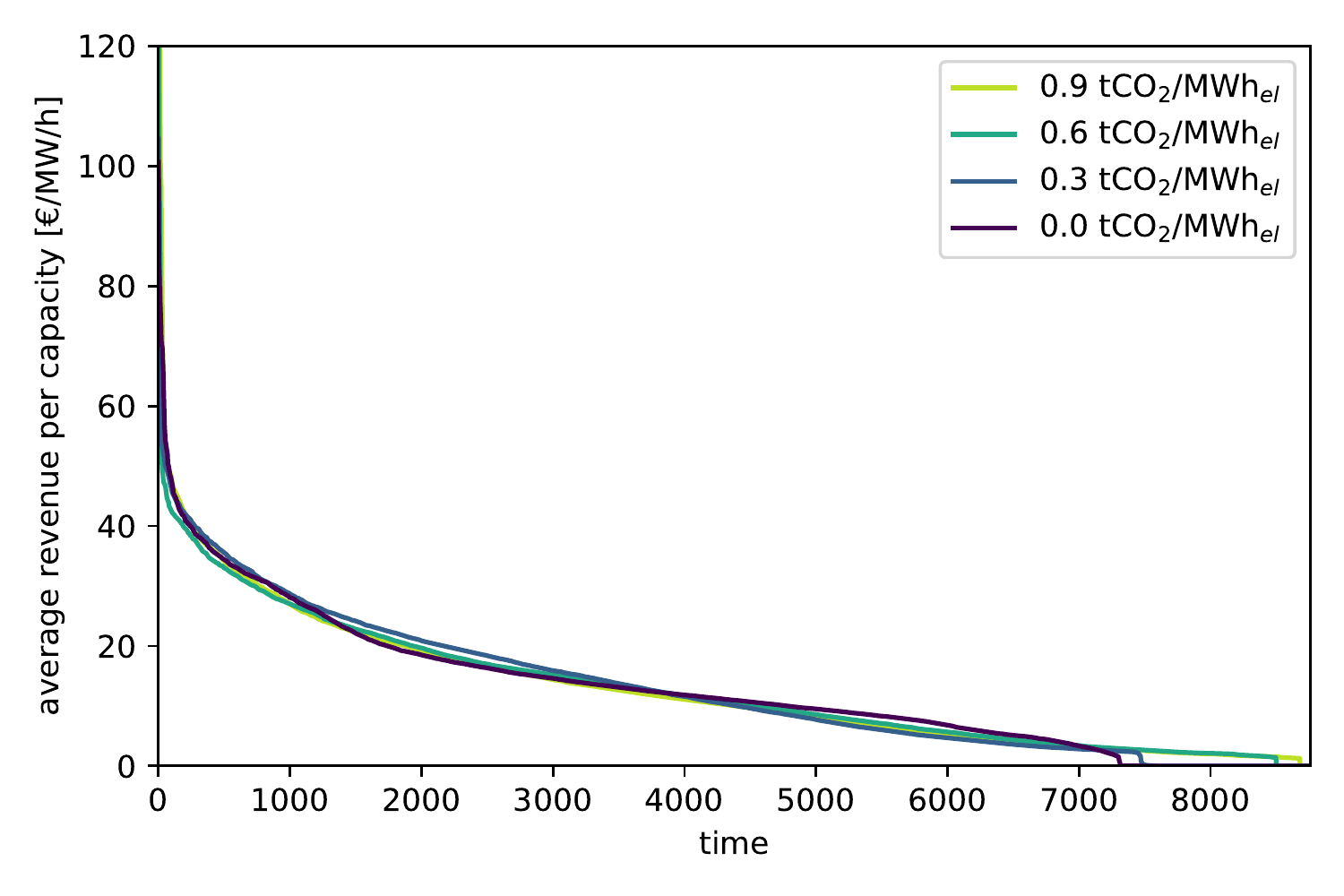}
\caption{Average hourly per-capacity revenue duration curve for wind for different average system CO$_2$ emissions, with flexibility from storage and transmission reinforcement.}
\label{fig:average_revenue-co2-wind}
\end{figure}

\begin{figure}[!t]
\centering
    \includegraphics[trim=0 0cm 0 0cm,width=\linewidth,clip=true]{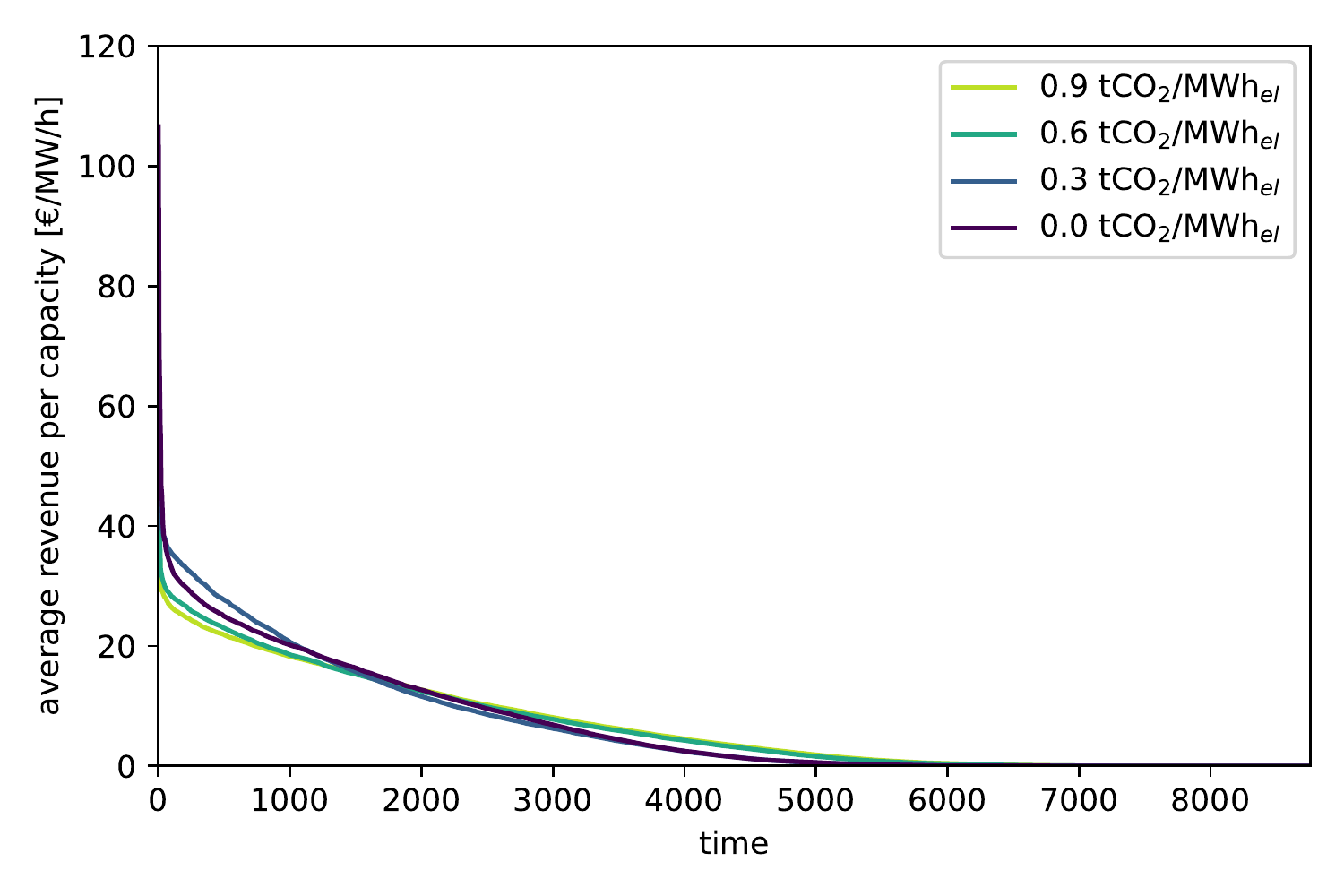}
\caption{Average hourly per-capacity revenue duration curve for solar for different average system CO$_2$ emissions, with flexibility from storage and transmission reinforcement.}
\label{fig:average_revenue-co2-solar}
\end{figure}

The distribution of average revenue per capacity for wind and solar over the hours is largely unchanged, see the duration curves in Figures \ref{fig:average_revenue-co2-wind} and \ref{fig:average_revenue-co2-solar}. The number of hours where the generators make their money shows no concentration into a few hours as emissions reduce, but is spread evenly through the year in all cases. Note that this is not the same as the market value; the area under the curve is instead equal to the annualised costs of wind and solar.

%~/energy/playground/market-value/duration_curves.ipynb

\bibliographystyle{elsarticle-num}

\biboptions{sort&compress}
\bibliography{market_value}

\end{document}